# Analytical Modeling the Multi-Core Shared Cache Behavior with Considerations of Data-Sharing and Coherence

Ming Ling, Xiaoqian Lu, Guangmin Wang, Jiancong Ge
*National ASIC System Engineering Technology Research Center, Southeast University, Nanjing, China*
{trio, lxqian, 220174427, gejiancong }@seu.edu.cn

***Abstract*—To mitigate the ever worsening "Power wall" and "Memory wall" problems, multi-core architectures with multi-level cache hierarchies have been widely accepted in modern processors. However, the complexity of the architectures makes modeling of shared caches extremely complex. In this paper, we propose a data-sharing aware analytical model for estimating the miss rates of the downstream shared cache under multi-core scenarios. Moreover, the proposed model can also be integrated with upstream cache analytical models with the consideration of multi-core private cache coherent effects. This integration avoids time-consuming full simulations of the cache architecture that required by conventional approaches. We validate our analytical model against gem5 simulation results under 13 applications from PARSEC 2.1 benchmark suites. Compared to the results from gem5 simulations under 8 hardware configurations including dual-core and quad-core architectures, the average absolute error of the predicted shared L2 cache miss rates is less than 2% for all configurations. After integrated with the refined upstream model with coherence misses, the overall average absolute error in 4 hardware configurations is degraded to 8.03% due to the error accumulations. The proposed coherence model can achieve similar accuracies of state-of-the-art approach with only one tenth time overhead. As an application case of the integrated model, we also evaluate the miss rates of 57 different multi-core and multi-level cache configurations.**

## I. INTRODUCTION

Performance evaluation plays an important role in the design cycle of the next generation processors as it allows architects to choose the architectural parameters for optimal performance and energy consumption trade off. Earlier researchers use cycle-accurate simulations to evaluate designs for their high accuracies [1]. However, these simulations are extremely time-consuming due to the ever-growing complexities and scales of architecture design spaces and workloads. Therefore, in the early stage of design cycles, architects prefer to use analytical models for their higher efficiency. Moreover, analytical models provide more insights that enable us to trade off different performance parameters in the architecture design.

To model the cache miss rates, which are critical for the processor performance evaluations, most analytical models take the cache configuration parameters and locality metrics profiled from the memory accessing stream as their inputs, such as the Stack Distance Histogram (SDH) and the Reuse Distance Histogram (RDH). Combined with the mechanistic analyses and probability derivations, these models can estimate the cache miss rates. However, when applied in the lower level caches, e.g., L2 or L3 caches, this approach requires profiling the memory reference streams to the target cache levels instead of the memory references directly from the CPUs, which can be normally obtained by a binary instrumentation tool in a relatively low overhead. In this case, they need time-consuming simulations of the cache architecture to profile the memory accessing streams to the target cache level [2][3][4]. For example, the reference stream to the L2 cache shared by the multi-core, shown in Fig. 1[1], consists of the streams from each individual core. To obtain the RDH/SDH of the merged shared cache reference stream, these previous works have to extract the individual SDHs/RDHs from each core's private cache to the shared L2 cache from detailed simulations, which, to a large extent, nullifies the evaluation speed benefits of analytical modeling.

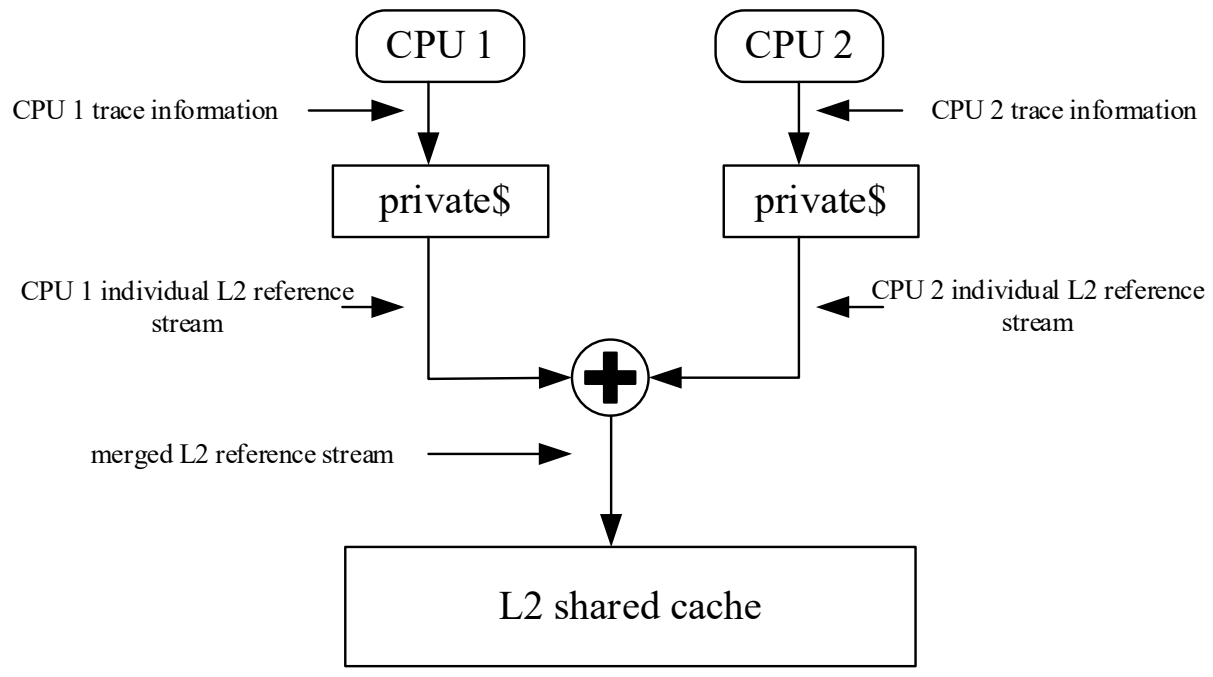

Fig. 1. An example of a multi-core processor with a multi-level cache hierarchy

Another factor that needs to consider is the data sharing among the threads running on different cores, i.e., different cores accessing L2 cache with same addresses. Some prior models, for example, StatCC [2], ignore this effect and evaluate the L2 behaviors merely based on the individual RDHs, which inevitably causes larger errors in multi-thread programs with intensive data sharing. Fig. 2 shows the result of the L2 miss rates of PARSEC 2.1[5] estimated by StatCC in a dual-core processor equipped with ALPHA ISA and the traditional 7-stage out-of-order pipeline. The cache hardware configuration can be found in the title of the figure. Because L2 cache is equipped with LRU replacement policy, StatStack [8] is used to calculate the miss rates of the L2 cache. Some data-sharing-intensive benchmarks, such as *canneal* and *vips*, have nearly 10% errors, which far exceeds the 2% average error in other benchmarks. Although Jiang's work [3] and Jasmine's work [4] quantify the data sharing effects in the L2 shared cache modeling, like we introduced above, their models require the L2 individual access streams that obtained from time-consuming simulations. In our previous work [9], a data-sharing aware L2 cache model that does not need full simulations as input has been proposed. However, neither the details of integrating the proposed model with the upstream cache mode nor the overall error evaluations after the integration have been discussed.

[1] Fig. 1 only shows a dual-core scenario. However, other multi-core configurations in this paper are equipped with the similar cache architecture.

Except for the data-sharing effect on the behavior of shared caches, we should also consider the coherence misses in the private caches under multi-core environment. A coherence miss occurs when one core tries to access a private cache line that has already been invalidated by the coherent protocol because another core changed the content of its private cache line with the same corresponding address. Coherence misses play a unignorable role in the performance evaluations of a multi-core processor [21][22]. Unfortunately, like many prior studies, the upstream cache model [6][7] used in this paper does not consider coherence misses for it was originally designed for a single-core environment.

In this paper, we propose a data-sharing aware shared cache miss rates model for multi-core systems with multi-level cache hierarchies. To eliminate the time-consuming full simulations, we integrate the proposed model with the upstream model [6][7] that put forward by our previous work, which outputs the individual L2 accessing RDHs from each core's private cache, to construct a multi-core multi-level cache model framework. The details of model integration and the overall error analysis after the integration have been introduced and discussed. Moreover, by using a similar analyzing method in the data-sharing aware model, we refine the upstream model to quantify coherence misses in each upstream private cache with similar accuracies and only one tenth time overhead compared to those of state-of-the-art approach [24].

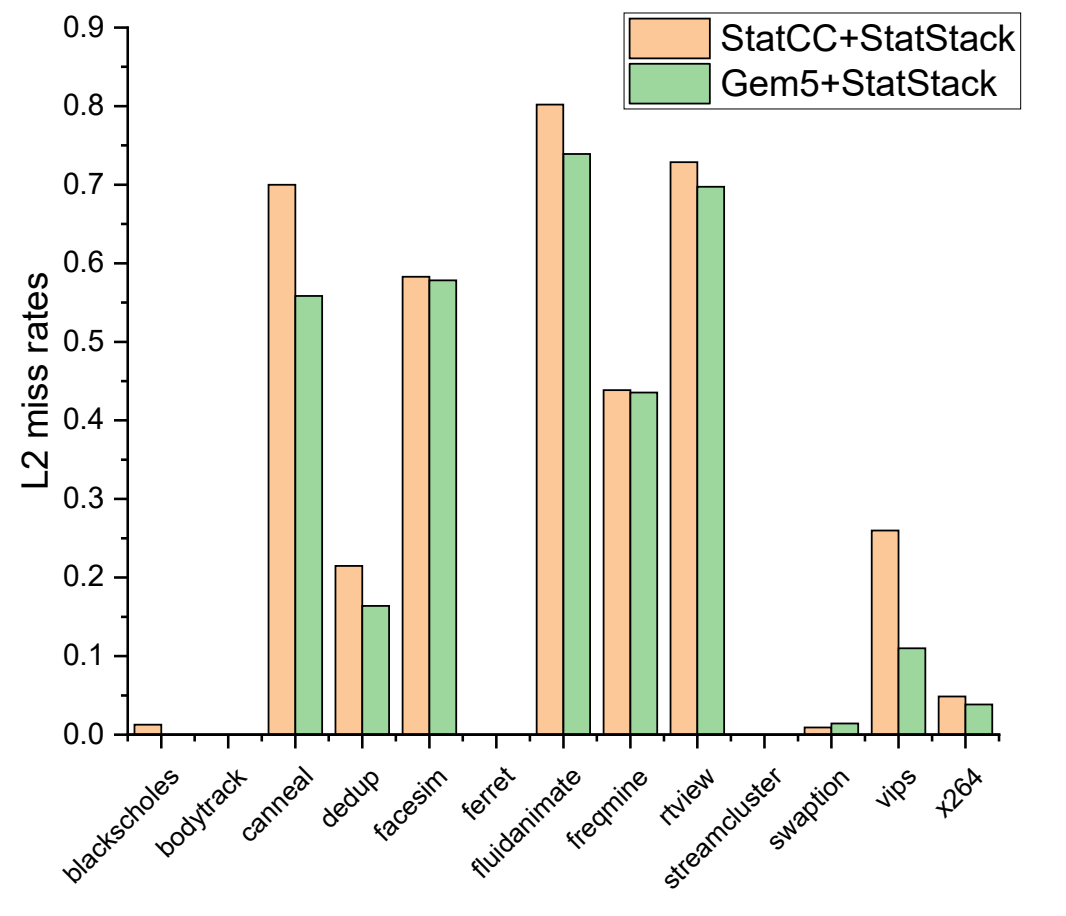

Fig. 2. The L2 miss rates estimated by StatCC for PARSEC(dual-core; L1 32KB; L2 2MB; DRAM 4GB; LRU-LRU)

The overview of the model framework is shown in Fig. 3. Firstly, Ge's model [6] [7] is modified and used as the upstream model to obtain each core's L2 individual Address Access Distributions, or AAD defined in section III, and individual RDHs accessing the L2 shared cache. Secondly, with considering of the coherence misses, we refine the private cache misses in the result of the first step. Thirdly, based on the individual RDHs and AADs from Ge's model, we construct locality information of the L2 shared cache MRDH, which also will be introduced in section III. Lastly, StatStack [8] is applied to calculate the miss rates based on the L2 shared cache MRDH obtained by our model.

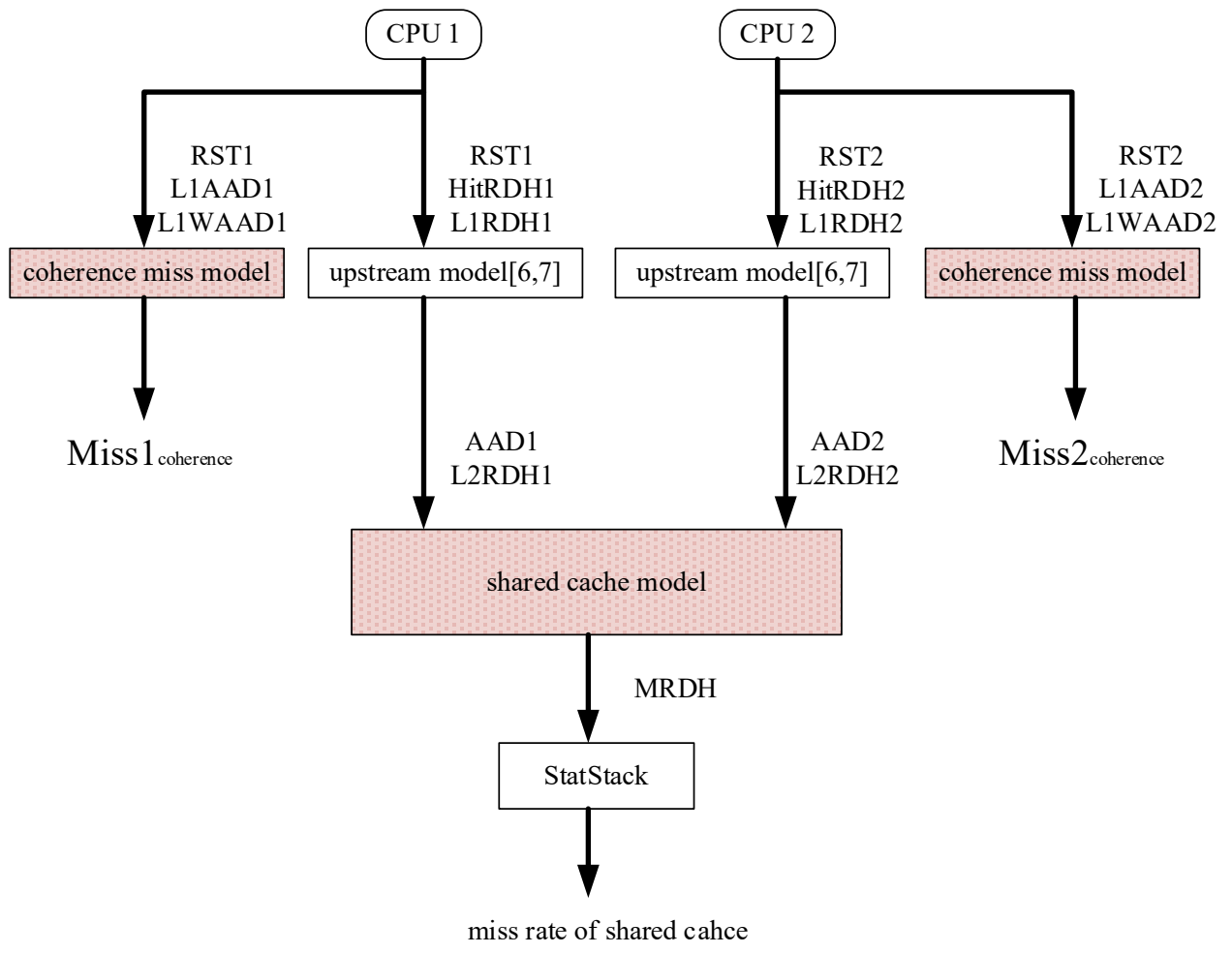

Fig. 3. The overview of the muti-level cache model framework

The main contributions of this paper can be summarized in the following three aspects:

- Providing an analytical method to quantify the influence of data sharing in the shared L2 cache.
- Thanks to its scalability, we combine our model with the upstream cache analytical model [6][7] to avoid the time-consuming multi-level cache simulations that required by prior approaches. The overall average errors of the integrated models have also been evaluated and discussed.
- Quantifying the coherence misses in the private caches, which enables the upstream cache model [6][7] be used in a multi-core scenario with a significant lower time cost.

The rest of the paper is organized as follows: Sections II introduces the related works. Section III introduces how our model constructs MRDH from individual RDHs with the consideration of data sharing. Section IV introduces how to integrate the shared cache model and the upstream cache model with the consideration of coherence misses. The evaluation results of our model are exhibited in section V. Section VI gives an application case of the proposed integrated model framework. Section VII concludes this paper.

## II. Related Works

The previous works in cache modeling can be categorized into three parts. The first part is for the models that focus on one certain cache-level in a uni-core processor. Erick Berg et al. [10] presented an analytical model, named StatCache, to estimate the L1 cache misses with the Random replacement policy. David Eklov et al. [8] proposed StatStack which derives SDH from RDH to predict L1LRU cache misses. X Pan et al. [11] utilized the Markov chain to predict the cache misses under three different replacement policies. For out-of-order processors, K Ji et al. [12] used artificial neural networks to address the effects of the stack distance migration caused by out-of-order executions.

The second part is for the models for multi-level caches in uni-core processors. K Ji et al. [13] [14] constructed a probability formula set to predict the downstream cache misses using the L1 cache SDH. To simplify the complexity of K Ji's algorithm, M Ling and J Ge [6] [7] proposed the RST table and Hit-RDH, which describe more detailed information of the software traces, as the inputs to model the L2 cache RDH. Jasmine Madonna S et al. [15] put forward an analytical model to calculate the L2 cache miss rate based on the analysis of the influence of cache inclusion/exclusion policies.

The third part is for the models with multi-level caches in multi-core processors. David Eklov et al. [2] proposed StatCC, a simple yet efficient model, which estimates the shared cache miss rates of co-scheduled applications on architectures similar to Fig. 1. It took the statistical locality characteristics of the memory reference streams in each core. However, it ignores the effects of data sharing among different threads, which causes larger errors when estimating multi-thread programs with intensive data sharing. Y Jiang et al. [3] provided a probabilistic model to find the merged stack distance profiled from the locality information of two individual threads. The input of their model is the SDH of each thread, which is obtained from time-consuming simulations of the upstream caches. Jasmine et al. [4] used Markov chains with combinatories and a basic probability theory to model the MSDH of multi-thread applications. Similar to Jiang's work, the model requires upstream cache simulations to obtain the inputs. Moreover, Chen Ding et al. [16] proposed a footprint theory based on the concept of the memory footprint. The theory also deals with the properties of footprint composition and proposes optimal co-scheduling by using shared footprint metric. Derek L. et al. [25] extended the traditional reuse distance theory on the multi-core platform to take invalidation and cache misses into account by capturing reuse distance consistently. M Wu et al. [26] raised up a model to analyze the changes of reuse distance by catching the profiling information of concurrent reuse distance in Loop-based Parallelism. For multicore GPU performance modeling, reuse distance cannot be directly applied for data interaction in multi-thread. Nugteren et al.[27] verified that the reuse distance theory can be used in GPU's thread by modeling caches in details approaching to the real-hardware application. Kiani et al. [28] took a trace-driven method which employed two reuse distance analysis (RDA) algorithms namely HLRDA and DRDA to evaluate the influence of hardware configurations in RDA. However, this model is just applied to the GPUs with the same hardware configurations.

Except for cold misses, conflict misses and capacity misses, there are also coherence misses in multi-core processors too. However, studies introduced above didn't consider the coherence misses, which significantly influence the performance of cache systems [21][22]. Berg et.al [23] presented a sample-based statistical model called StatCacheMP built from processors with multiple cores to analyze the data locality only with the Random replacement policy. To count the coherence misses, it has to monitor the reuse distance continuously and maintain a writer list which records the IDs of other cores writing to the monitored cachelines. If the writer list is not empty, the cachelines can be seemed as invalid and triggers coherence misses. However, these models use PIN[29] tool, which only supports X86 architectures, to catch traces of memory instructions. Furthermore, both of these models need to profile and analyze the memory accessing traces in an extremely time-consuming manner, which significantly degrades the evaluation efficiency.

In our previous work [9], we put forward a new model that considers the insertion effects and split effects of shared data accessing to quantify the merged L2 RDH, which can be used to calculate the L2 miss rates. To avoid the time-consuming cache simulations, it was declared in that paper that the proposed data-sharing aware downstream cache model can be integrated with the upstream cache model [6][7]. However, neither the integration details nor the error analysis after the integration have been introduced and discussed. Meanwhile, the upstream model in that work does not consider the coherence misses in a multi-core environment.

In this paper, we introduce the data-sharing aware shared cache model with the details of integrating the upstream cache model, such as the AAD extraction method. The error analysis after the integration have also been evaluated and discussed. We have also refined the upstream model to quantify the coherence misses in the private caches using a similar analyzing method in the data-sharing aware shared cache model with much less time overhead.

## III. Model MRDH from Individual RDHs

Before introducing our model, we first define some basic terminologies used in our following discussions.

*Reuse Distance*: The reuse distance is the number of references between two consecutive references accessing to the same cache line.

*Merged Reuse Distance*: The merged reuse distance is the reuse distance of the references in the merged reference stream to the shared cache, which is constructed by the interleaving of the individual reference streams from all cores in a multi-core system.

*Merged Reuse Distance Histogram (MRDH)*: MRDH records the numbers of references for every merged reuse distance in the memory traces in a given profiling interval.

*Access Address Distribution (AAD)*: AAD records the numbers of references to each cache-line aligned address in a given profiling interval.

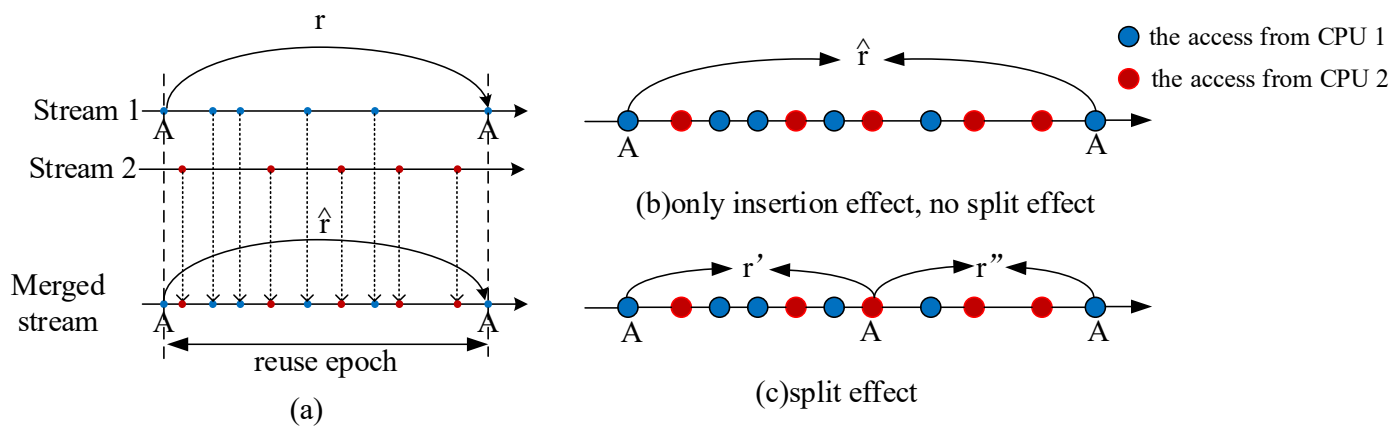

Fig. 4. The insertion effect and the split effect of accessing stream interleaving

To simplify the discussion, the model construction of a dual core architecture is taken as an example for the derivation. As shown in Fig.3, the inputs of the model are the L2 individual RDHs and individual AAD from each core, which are obtained by Ge's model[6][7]. In this case, we maintain an access address distribution table $AAD_i[addr_x]$, in which $addr_x$ is the cache line aligned address of the coming reference. The range of $x$ is from 1 to the total number of addresses accessed in the profiling interval, denoted as N in this paper. When a coming reference $x$ is an L1 miss, which means the reference will be leaked to the L2 cache, we will accumulate the element $AAD_i[x]$ by one. The subscript $i$ represents the reference comes from the *i-th* core.

As shown in Fig. 4(a), two reference streams from two cores interleave in the L2 shared cache and construct the merged stream. The interleaving of the two individual reference streams may change the reuse distance of the reuse epoch that constructed by two consecutive references A (i.e., the reference to the address A). We divide the changes of RDHs caused by the interleaving into two categories: 1) As shown in Fig. 4(b), the reuse distance of the reference A increases because of the references from the other core, which is named as the insertion effect; 2) As shown in Fig. 4(c), when the address of an inserted reference is same as the endpoint of the reuse epoch, i.e., A in this case, the original reuse epoch is split into two new reuse epochs, which is called the split effect.

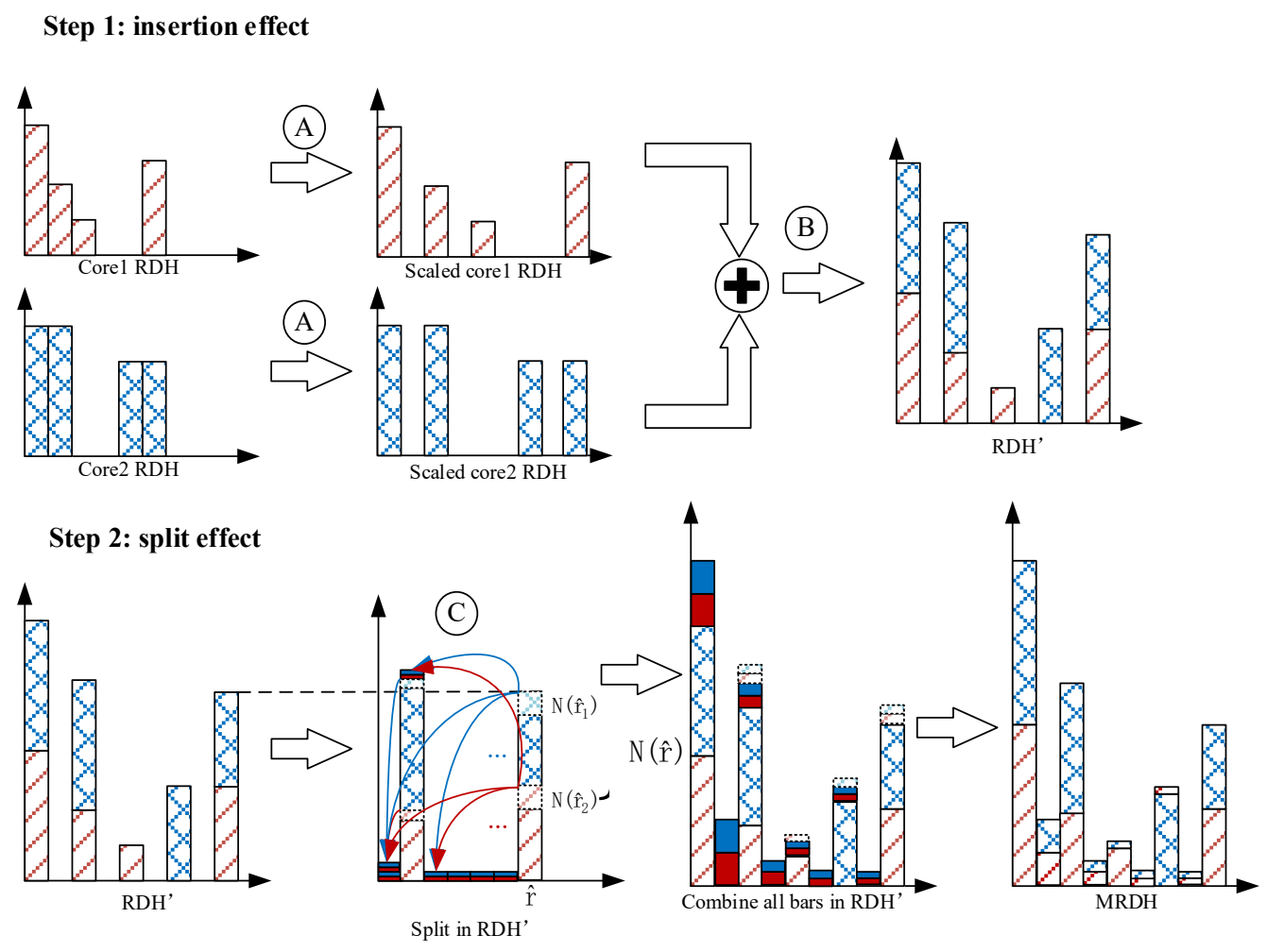

Fig. 5. The two steps of our work

According to these two effects, we construct the model in two steps, shown in Fig. 5: 1) Quantifying the insertion effect caused by the multi-core reference stream interleaving; 2) Quantifying the split effect on reuse epochs to refine the result of the first step with data sharing.

## *A. Quantifying the insertion effect*

In this step, we use an approach similar to StatCC to quantify the insertion effect. The reuse distance of the reference from one core is stretched because of the insertion of references coming from another core. For example, as shown in Fig. 4(b), the reuse distance of the reference A in Core 1 is 4(blue dots), while the reference stream of Core 2 inserts 5 references (red dots) into the reuse epoch. Thus, the reuse distance of the reference A is stretched to 9 with the scale of $^9/_4$. If the ratio of the number of accesses from different cores in an interval remains relatively uniform and stable, the scale of the stretch can be regarded as a constant for all reuse epochs. By multiplying the original reuse distance with this constant, we can calculate the merged reuse distance after the streams being interleaved. The derivation is as follows:

For a reuse epoch from Core 1 with a reuse distance of $r_1$, Core 2 also accesses the shared cache during the period of the reuse epoch. Supposing that Core 2 generates $r_2$ references during this period, $r_1$ will be stretched to $\hat{r}_1$ as shown in Eq. (1).

$$\hat{r}_1 = r_1 + r_2 \tag{1}$$

Similar to StatCC, we assume that the shared cache accesses are uniform in the whole profiling interval, i.e., we do not consider the effect of program phase transitions. Therefore, the relative speeds of Core 1 and Core 2 accessing the shared cache are unchanged in the interval. In the entire profiling interval, Core 1 accesses the shared cache $access_1$ times, and at the same time Core 2 accesses the shared cache $access_2$ times, from which we can derive the approximate relationship as shown in Eq. (2).

$$\frac{access_2}{access_1} = \frac{r_2}{r_1} \tag{2}$$

The total number of references $access_1$ from Core 1 and $access_2$ from Core 2 can be calculated by Eq. (3), where N denotes the number of different addresses during this profiling interval.

$$access_i = \sum_{x\in[1,N]} AAD_i[addr_x] \tag{3}$$

Bring Eq. (2) to Eq. (1), we get the following relationship as Eq. (4):

$$\hat{r}_1 = r_1\left(1 + \frac{access_2}{access_1}\right) \tag{4}$$

The analysis of Core 2 is similar, so we get Eq. (5).

$$\hat{r}_2 = r_2\left(1 + \frac{access_1}{access_2}\right) \tag{5}$$

The stretch of the reuse distance obtained by the insertion effect can be quantified by Eq. (4) and Eq. (5), which reflect the procedure (A) in Fig. 5. According to the scale of the stretch, the insertion effect can be described by Eq. (6):

$$\begin{cases} RDH'(\hat{r}) = RDH1(r_1) + RDH2(r_2) \\ r_1 = \hat{r}\Big/\left(1 + \dfrac{access_2}{access_1}\right) \\ r_2 = \hat{r}\Big/\left(1 + \dfrac{access_1}{access_2}\right) \end{cases} \tag{6}$$

In Eq. (6), $\hat{r}$ denotes the merged reuse distance. According to the core that the reference comes from, $\hat{r}$ can be specified as $\hat{r}_1$ or $\hat{r}_2$. As shown in Fig. 5 (A), we can calculate the $\hat{r}_1$ and $\hat{r}_2$ by applying Eq. (4) and Eq. (5) to all reuse epochs of Core 1 and Core 2, respectively. By combining all the calculation results as shown in Fig. 5 (B), a new RDH can be obtained, denoted as $RDH'$. The result $RDH'$ is a locality metric that quantifies the insertion effect without considering the split effect, or the influences caused by data sharing.

## *B. Quantifying the split effect*

In order to quantify the split effect, we first calculate the probability of a reuse epoch with $r$ memory references being split, denoted as $P_{split}(r)$.

As we shown in Fig.4(c), the inserted references from the other core may include one or more references that access the same address as the endpoints due to data sharing. Once that happens, the original reuse epoch will be split. In Fig. 4(c), the reuse epoch is constructed by address A, therefore we first estimate the probability of epochs constructed by address A among all the reuse epochs of Core 1.We assume that Core 1 generates $n2$ memory references, in which there are $n1$ references accessing address A. The probability of the reuse epochs constructed by address A can be estimated as $n1/n2$. Also, a split only occurs when a reference b accessing A, in which the reference b represents any reference coming from Core 2 in the reuse epoch. During the same time span, if Core 2 generates $n3$ references accessing address A and the number of the references accessing the same cache set as address A is $n4$, the probability of reference b accessing A can be estimated as $n3/n4$. It should be noted that we use the number of references accessing A's set in the calculation instead of the total number of references from Core 2 because only the reference accessing the A's set will be counted into the reuse epoch constructed by address A. In conclusion, the probability of the split in references with address A, $P1_{same-A}$, can be calculated as $(n1/n2) \times (n3/n4)$.

The total probability of any memory reference coming from Core 2 accessing same address as the endpoints is the sum of $P1_{same-addr_x}$, in which $addr_x$ denotes any shared address of the shared cache accessed by Core 1 and Core 2. Therefore, the derivation in common scenarios can be represented as Eq. (7).

$$P1_{same} = \sum_{addr_x \epsilon S} \left(P1_{same-addr_x}\right)$$
$$= \sum_{addr_x \epsilon S} \left(\frac{AAD_1[addr_x]}{access_1} \times \frac{AAD_2[addr_x]}{\sum_{y \in x' set} AAD_2[y]}\right) \quad (7)$$

In Eq. (7), $S$ represents the shared address set of the shared cache accessed by Core 1 and Core 2. $AAD_1[addr_x]/access_1$ means the ratio of references accessing $addr_x$ in all the L2 shared cache references from Core 1. To split the reuse epochs of $addr_x$, the coming references from another core should access the same address, whose probability can be described as $\frac{AAD_2[addr_x]}{\sum_{y \in x' set} AAD_2[y]}$. The numerator $AAD_2[addr_x]$ is the number of the references that accessing the same address $addr_x$and the denominator $\sum_{y \in x' set} AAD_2[y]$ is the number of the Core 2 references that accessing the same set with $addr_x$ during this profiling interval. The ratio means the probability of a reference coming from Core 2 accessing the same address $addr_x$ in all references from Core 2 that accessing the same cache set with $addr_x$.

Similarly, the corresponding probability for Core 2 can also be calculated in Eq. (8).

$$P2_{same} = \sum_{addr_x \epsilon S} \left(\frac{AAD_2[addr_x]}{access_2} \times \frac{AAD_1[addr_x]}{\sum_{y \in x' set} AAD_1[y]}\right) \quad (8)$$

Eq. (7) and (8) describe the probability of any reference in a reuse epoch accessing the same address as the endpoint reference. The reuse epoch will be split if there exist one or more inserted memory references in a reuse epoch accessing the same address as the endpoint. As every inserted reference coming from another core has the probability $P_{same}$ to accessing same address with the target core, the probability of the reuse epochs with the reuse distance $r$ being spilt, denoted as $P_{split}(r)$, can be calculated by subtracting the not splitting probability from 1, represented as Eq. (9). Note that $r$ is the reuse distance before the stream merging, while the reuse distance after merging is denoted as $\hat{r}$. Therefore, the number of inserted memory accesses form the other core is $\hat{r} - r$.

$$P_{split}(r) = 1 - (1 - P_{same})^{\hat{r}-r} \quad (9)$$

Eq. (9) can be specified for Core 1 and Core 2 as shown in Eq. (10):

$$\begin{cases} P_{split}(r_1) = 1 - (1 - P1_{same})^{\hat{r}_1 - r_1} \\ P_{split}(r_2) = 1 - (1 - P2_{same})^{\hat{r}_2 - r_2} \end{cases} \quad (10)$$

Based on Eq. (10), the number of reuse epochs splitting on the bar $\hat{r}$ in $RDH'$, denoted as $N(\hat{r})$, can be calculated as Eq. (11).

$$\begin{cases} N(\hat{r}) = RDH1(r_1) \times P_{split}(r_1) + RDH2(r_2) \times P_{split}(r_2) \\ r_1 = \hat{r} \Big/ \left(1 + \dfrac{access_2}{access_1}\right) \\ r_2 = \hat{r} \Big/ \left(1 + \dfrac{access_1}{access_2}\right) \end{cases} \quad (11)$$

Eq. (12) gives the way to calculate the merged reuse distance histogram from both cores *MRDH($\hat{r}$)*. In this equation, $RDH'(\hat{r})$ is predicted in Step 1 of Fig. 5. Because of the split effect, the reuse distance of the references may be decreased, i.e., some references with high reuse distance will be migrated to the bars with lower reuse distances in Fig. 5. Thus, $N(\hat{r})$ means the number of references with the original reuse distance of $\hat{r}$ that will be spread to the bars with lower reuse distances, which illustrated as dash boxes in Fig. 5 (C). Assuming that the references evenly migrated to the bars with lower reuse distances, $\sum_{rd=\hat{r}+1}^{\infty} \frac{N(rd)}{rd}$ in Eq. (12) means the reference numbers migrated from the bars with higher reuse distances to the $\hat{r}$ bar, which are represented as red boxes and blue boxes shown in Fig. 5 (C).

$$MRDH(\hat{r}) = RDH'(\hat{r}) - N(\hat{r}) + \sum_{rd=\hat{r}+1}^{\infty} \frac{N(rd)}{rd} \quad (12)$$

Fig. 6 gives the methodology of extending our model into a quad-core scenario. In this figure, there are 4 cores connected to the L2 shared cache. When we predict the portion of MRDH contributed by Core 1 (target core), we consider the other three cores as a black box, called virtual Core_v2 in Fig. 6 and the reference streams from other three cores are just considered from Core_v2. Therefore, when we calculate $P_{same}$ for Core 1, the probability that the coming reference from other three cores accessing the same address with the endpoint reference will be described as $\frac{AAD_{core_v2}[addr_x]}{\sum_{y \in x' set} AAD_{core_v2}[y]}$. After considering each core as the target core and accumulating the portions of MRDH contributed by all the 4 cores, MRDH of the L2 shared cache can be obtained. Processors with more cores can be predicted in the same way.

## IV. INTEGRATING WITH THE UPSTREAM CACHE MODEL

The proposed shared cache model can be integrated with the upstream cache model [7], which provides AAD and RDH to the shared cache model, to avoid time-consuming simulations. However, before the integration, there are still two problems need to solve: 1) Extracting AAD by adding customized code in the original upstream cache model; 2) Quantifying the coherence misses that ignored by the original model. We will start from the introduction of the upstream cache model and how to extract needed information.

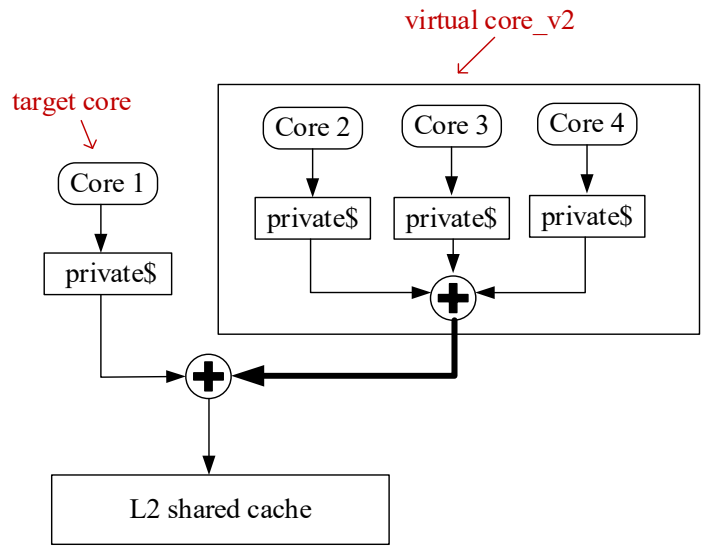

Fig. 6. Extending our model into a quad-core environment

## A. Upstream cache model

In this paper, we choose Ge's multi-level cache model [6][7] as the upstream cache model. The multi-level cache model proposes two new metrics, namely the Reuse-and-Stack-Transfer (RST) table and the Hit-RDH table. RST table is a two-dimensional matrix, which records information of the RDH and the SDH in a given trace profiling interval in the L1 cache. As the example shown in Fig. 7, every element in the RST table contains the relationship between the reuse distance and the stack distance. The red circle $RST[4][1]$ in Fig. 7 represents there are 320 references in this interval with the reuse distance of 4 and the stack distance of 1. Moreover, for given references with the reuse distance of $i$, we use Eq. (13) to calculate the normalized RST table, called $P_{rs}$. This model defines each element $P_{rs}[i][j]$ as the probability that the references have the stack distance of $j$, which is the proportion of $RST[i][j]$ in the whole $i^{th}$ row. For instance, as shown in Fig. 7, the red circle in the normalized RST table means that in all references with the reuse distance of 4, 76% references have the stack distance of 1.

$$P_{rs}[i][j] = \frac{RST[i][j]}{\sum_{k=0}^{i} RST[i][k]} \qquad (13)$$

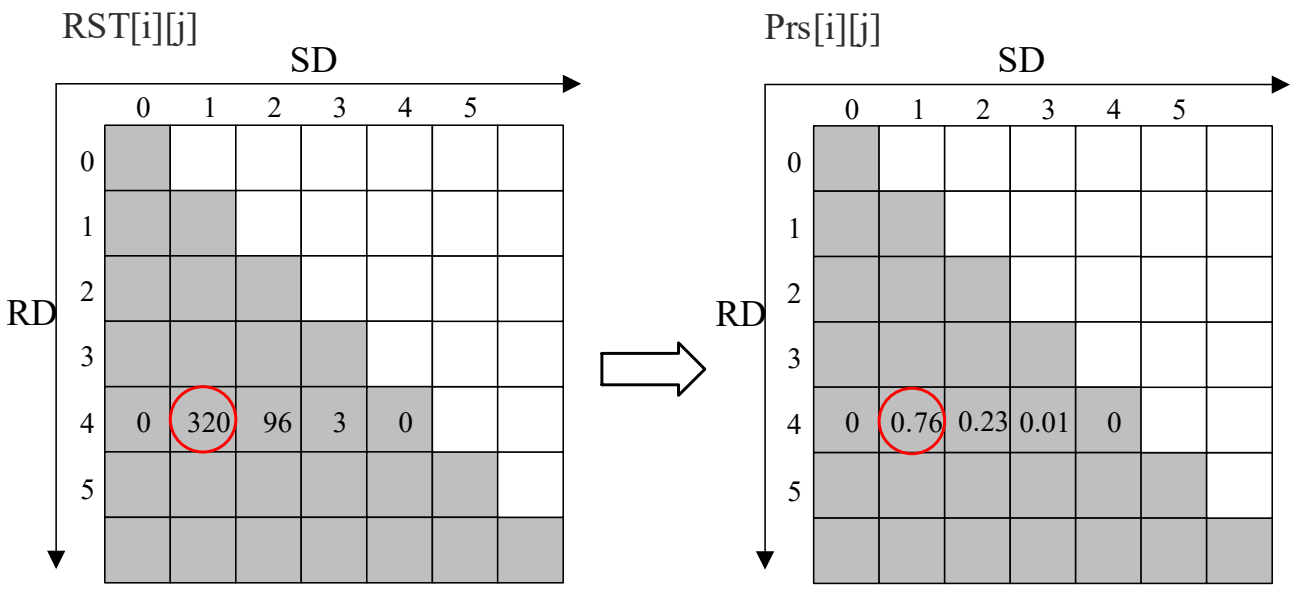

Fig. 7. The RST table and Normalized RST table ($P_{rs}$)

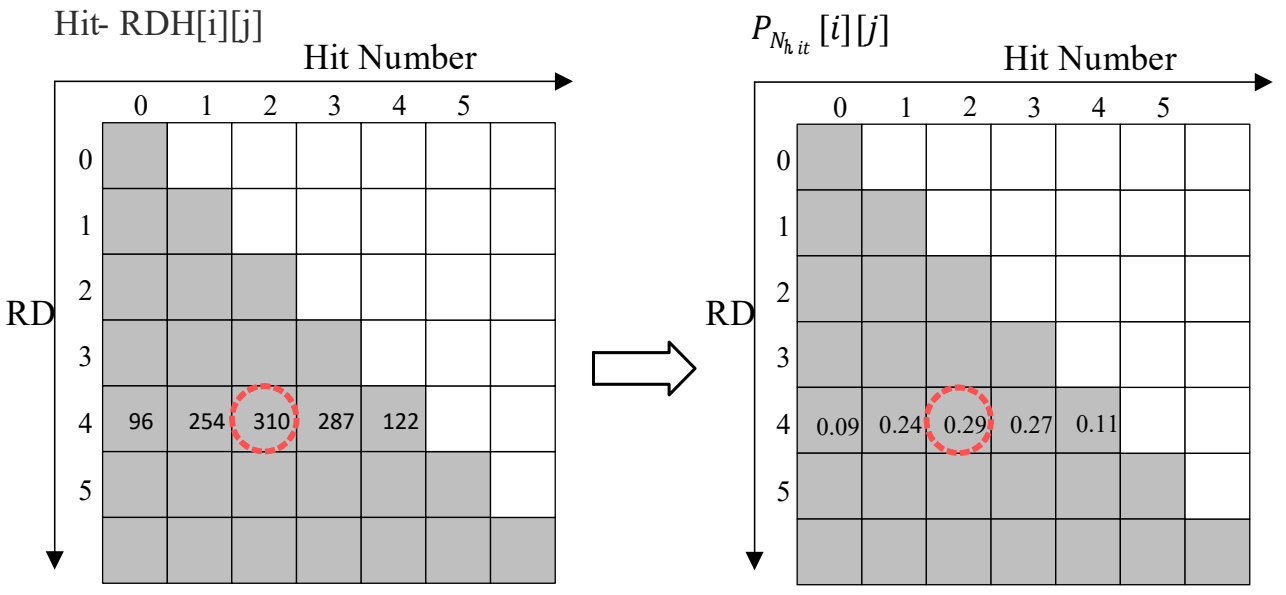

Fig. 8. The Hit-RDH and Normalized Hit-RDH ($P_{N_{hit}}$)

Another metric, Hit-RDH, introduced in the model is also a two-dimensional matrix. Fig. 8 shows an example of Hit-RDH. The red circle in Fig. 8 means that in all the reuse epochs with the reuse distance of 4, the number of reuse epochs that have 2 references hitting in the L1 cache is 310. In other words, there are 310 reuse epochs whose reuse distance are 4 and in each of them there are 2 references hit in the L1 cache. By Eq. (14), we can also get the normalized Hit-RDH, called $P_{N_{hit}}$, as shown in Fig. 8. $P_{N_{hit}}[i][j]$ is the proportion ofthe $Hit_{RDH}[i][j]$ in the whole $i^{th}$ row.

$$P_{N_{hit}}[rd][n] = \frac{Hit_{RDH[rd][n]}}{\sum_{k=0}^{rd} Hit_{RDH[rd][k]}} \qquad (14)$$

To construct RST and Hit-RDH tables for a set-associative cache, we need to maintain two linked-lists to record the reuse/stack history of each memory access that indexed to every individual cache set. When a memory reference A comes, as Fig. 9 shows, the index bits are firstly used to address the corresponding set linked-lists, while the extracted tag is pushed to ends of the reuse reference list and the stack reference list. By using this method, we can get the reuse distance and the stack distance for each memory reference. To construct the RST and Hit-RDH tables, we just need to increase the value by one in the corresponding element of each table for each coming reference, as shown in Fig. 9. Similarly, Fig. 9 also shows the process of obtaining AAD. After calculating the stack distance, we compare the stack distance with the associativity of the LRU L1 cache. If the stack distance is no less than associativity, which indicates a cache miss, we accumulate 1 on the corresponding memory address in AAD. To make a tradeoff between space/time overheads and accuracies, we choose to cut of the profiled L1 reuse/stack distances at 1024 and accumulate the number of references with larger reuse distances to the reuse distance bar of 1024, which is also applied in the work of [11, 12, 13, 16]. The RST, Hit-RDH and AAD updating procedures are just attached to the progress of RDH and SDH profiling. Thus, the extra time overhead of maintaining these three tables is negligible.

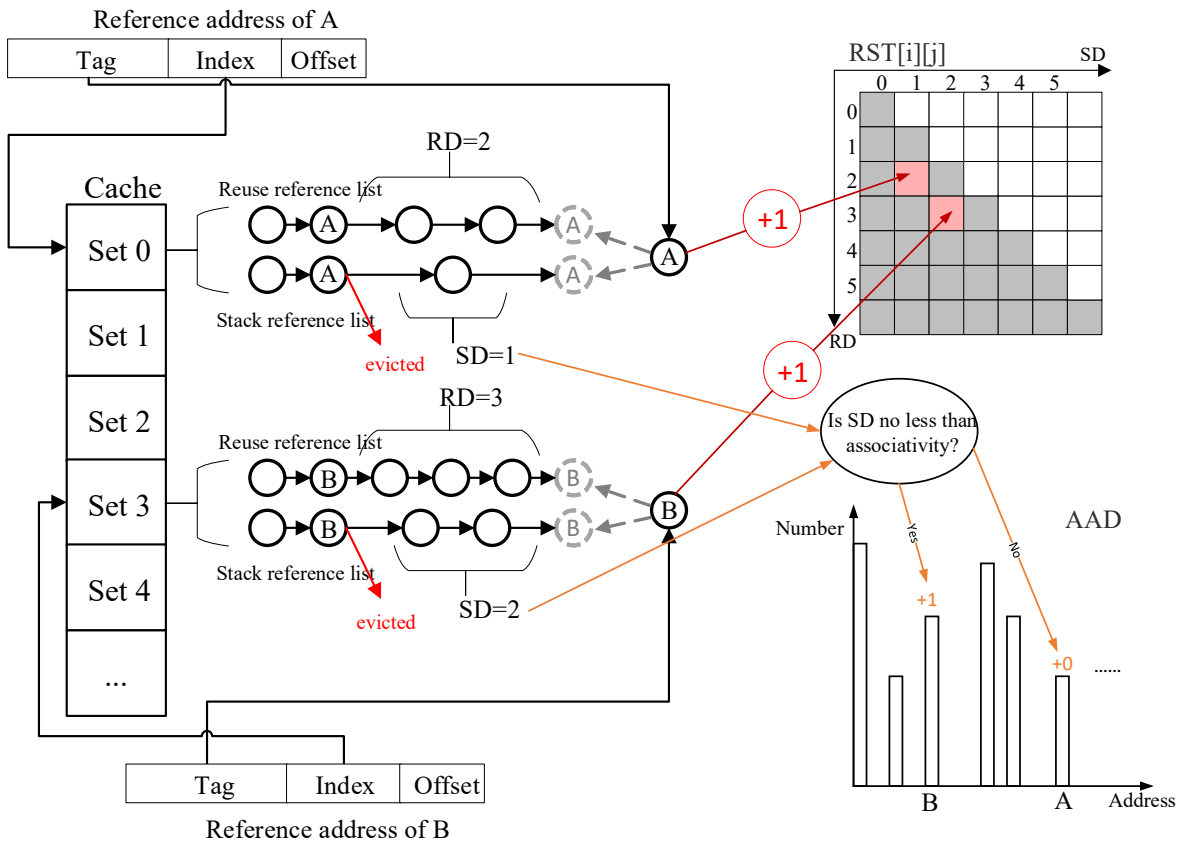

Fig. 9. TheReference lists used to extract RST and AAD

Limited by the space, we only introduce the key derivation of the upstream cache model, interested readers can refer to the original paper [7] to get more detailed information.

After filtering by L1 cache, the changes of L1 cache RDH can be divided into two parts: 1) Some references are hit in the L1 cache, so the L2 cache will not be accessed, and the total number of memory accesses to the L2 cache will be reduced. Reflected on the RDH, the height on the histogram will be decreased; 2) For a given reuse epoch in the L1 cache, some references between the epoch endpoints may hit in the L1 cache. Thus, when the reuse epoch leaked into the L2 cache (assuming the two endpoints are all misses), its reuse distance might be decreased (fewer references remained between the end points of the reuse epoch) and the reuse epoch should be counted to a lower L2 reuse distance. Fig. 10. shows these two steps for estimating L2RDH from L1RDH.

For a reference accessing to L1 cache, the model use $P_{rs}$ to estimate its stack distance. According to the definition of $P_{rs}$, $RDH(i) \times P_{rs}[i][j]$ represents the number of references that reuse distance is $i$ and stack distance is $j$ in the RDH. If stack distance $j$ is less than the associativity, these references will hit in L1 cache. Otherwise, they will access L2 cache and become parts of L2RDH. Therefore, the reduced number of references in RDH can be represented by Eq. (15).

$$MissRDH(i) = RDH(i) \times \left(1 - \sum_{j=0}^{L1\ Assoc-1} Prs[i][j]\right) \quad (15)$$

By the definition, $P_{N_{hit}}[rd][n]$ means that in all the reuse epochs with the reuse distance of $rd$, the proportion of the reuse epochs which have $n$ hit references in each of them. If the reuse distance of a reuse epoch in the L1 cache is $rd$ while its L2 reuse distance is $i$, this means there are $rd - i$ references in this reuse epoch hit in the L1 cache and the ratio of these references is $P_{N_{hit}}[rd][rd-i]$. Eq. (16) shows the way to get the L2RDH from $MissRDH$. In this equation, $MissRDH(rd) \times P_{N_{hit}}[rd][rd-i]$ represent show many memory references with L1 reuse distance $rd$ have been moved, or migrated, to a L2 reuse distance bar $i$ because there are averagely $rd - i$ references are L1 hits in each of the reuse epoch. By accumulating all the migrated references from each higher bar $(rd > i)$ in $MissRDH(rd)$, we can obtain the adjusted $L2RDH(i)$ as shown in Eq. (16).

$$L2RDH(i) = \sum_{rd=i}^{\infty} MissRDH(rd) \times P_{N_{hit}}[rd][rd-i] \quad (16)$$

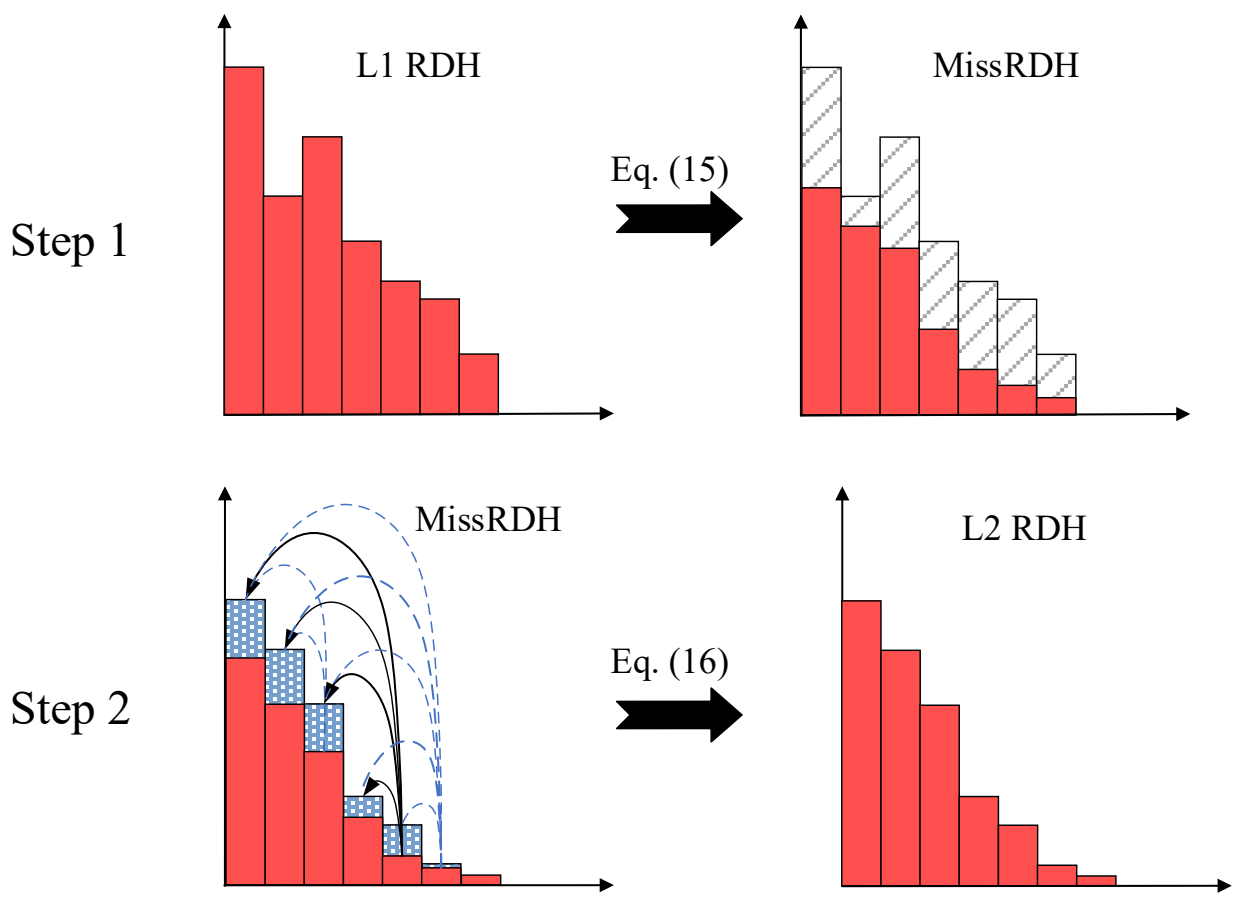

Fig. 10. Two steps from L1 RDH to L2 RDH

## B. *Quantifying private cache coherence misses*

Unfortunately, the introduced upstream cache model is built in a uni-core architecture without considering coherence misses in multi-core architectures. To integrate the model with the shared cache model, we must quantify the upstream private cache coherence misses.

We first make some assumptions to simplify our discussion:

- The coherent protocol is a protocol based on write invalid rather than a write update.
- When there is a coherence miss, the cache controller will obtain the cache line from other private caches or from the main memory without accessing the downstream cache. The reason why this assumption is needed is that the current write invalid coherent protocol, such as MESI (Modified, Exclusive, Shared, Invalid), only guarantee the state of a cache line without implementation details [18]. To simplify our model, we make this assumption.
- There are little coherence misses in the shared cache. Generally, coherence misses occur in the L1 private cache. The shared cache, e.g., L2 cache, is less affected by coherence misses than L1 private caches. Moreover, coherence misses of the shared cache are affected by write backs, which is hard to quantify by an analytical model based on RDH.
- Memory accesses are independent and uniformly distributed on different addresses.

Fig. 11 gives an example of a coherence miss. The shared cache line A is accessed by the two cores. At time t1, Core 1 accesses the cache line A, its state is set to E(Exclusive) in private cache of CPU 1. Then at the time t2, another core CPU 2 writes new data to cache line A and sends an invalid signal to CPU 1 to invalidate the cache line A in the private cache of CPU 1, the states of cache line A in private cache of CPU 1 and CPU 2 are I(Invalid) and M(Modified), respectively. Therefore, when the CPU 1 accesses the cache line A again at time t3, the cache line A is still in private cache of CPU 1 but its state is I(Invalid), which causes a coherence miss in the private cache of CPU 1.

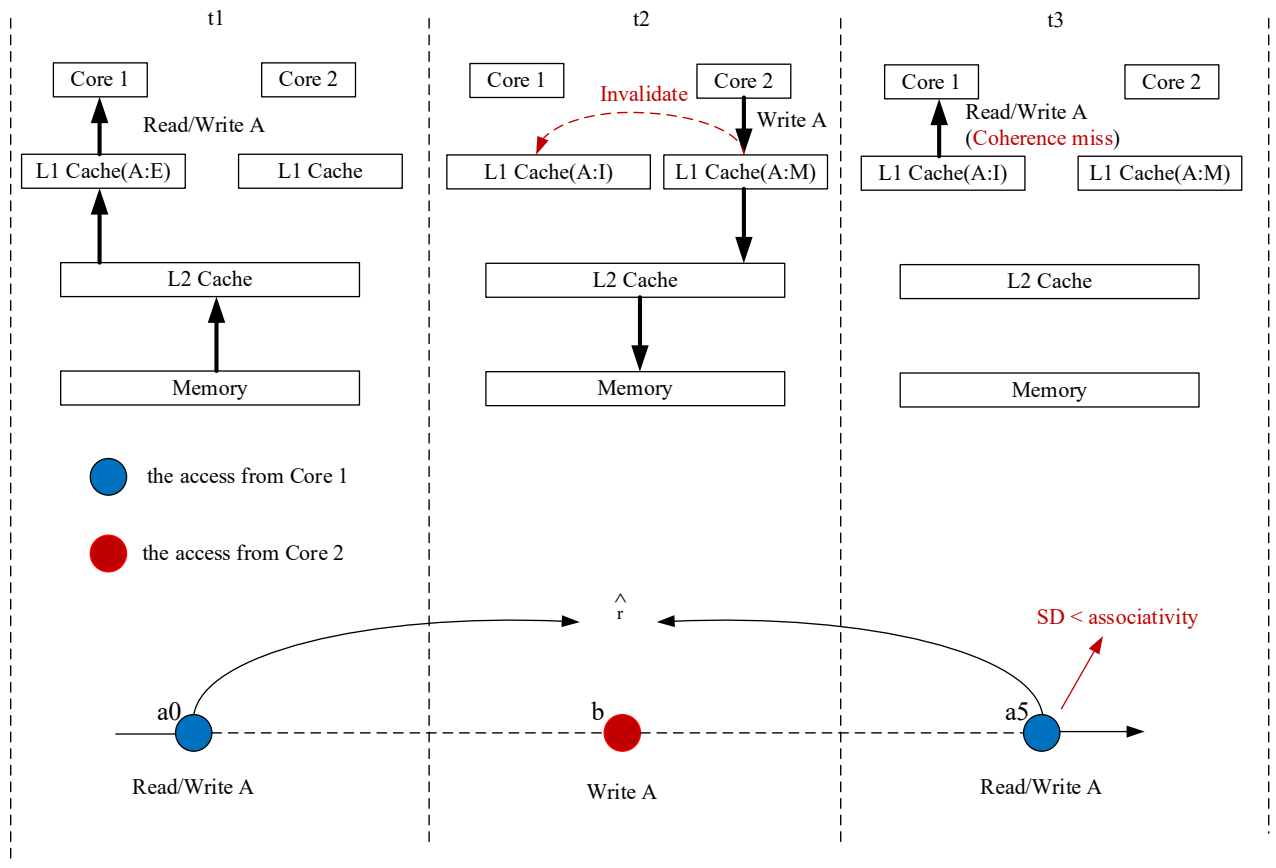

Fig. 11. An example of coherence miss

As shown in Fig. 11, the coherence miss occurs when a core accesses an invalid cache line in its private cache. The invalid state is caused by receiving invalid signals from other cores. The invalid signal is sent at the moment when the shared cache is written by other cores to change its content. Moreover, if the invalid cache line was evicted before the second reference, the miss is a capacity miss or a conflict miss instead of a coherence miss. Therefore, two conditions must be met when a coherence miss occurs: 1) During the period of two consecutive references from one core to an address, there are a write operation to the same address from another core; 2) When the second reference in the first core accesses the cache line, it is still in the private cache but with an invalid state.

As the coherence miss occurs after write references to a shared cache line, we need obtain extra two input parameters, L1AAD (L1 private cache Access Address Distribution containing write and read accesses) and L1WAAD (L1 private cache Write Access Address Distribution), to quantify the coherence miss. The definitions of these two parameters are similar to AAD described in Section III. The difference is that L1AAD and L1WAAD is the address distribution from CPU that accesses its L1 private cache, while AAD in Section III is the accessing address distribution from L1 caches to L2 shared cache. Both L1AAD and L1WAAD can be obtained directly from the CPU traces generated by a trace generator or a binary instrumentational tool.

As we have known, a coherence miss occurs after the write reference to the shared cache line. Therefore, the way of modeling coherence miss is similar to modeling data sharing. Thus, modeling the probability of coherence miss also can be divided to two steps: 1) Calculating the probability $P_{same-write}$ that any reference comes from another core accesses same address with the endpoint of the reuse epoch of the target core; 2) Calculating the probability $P_{split-write}$ that reuse epochs are split by shared write reference coming from another core. With the experience of modeling data sharing in Section III, we can analogize to calculate $P_{same-write}$ in Eq. (17) according to $P_{same}$.

$$P1_{same-write} = \sum_{addr_x \epsilon S} \left(P1_{same-write-addr_x}\right)$$
$$= \sum_{addr_x \epsilon S} \left(\frac{L1AAD_1[addr_x]}{L1_{access_1}} \times \frac{L1WAAD_2[addr_x]}{\sum_{y \in x' set} L1AAD_2[y]}\right) \quad (17)$$

In the Eq. (17), $S$ represents the shared data set accessed by the two cores, $addr_x$ is the address of the shared data in the set $S$. $L1AAD_1[addr_x]$ is the number of references that access $addr_x$ in L1 private cache of Core 1. $L1_access_1$ is the total number of the reference from Core 1. Therefore, their ratio $\frac{L1AAD_1[addr_x]}{L1_access_1}$ represents the probability that reuse epochs are constructed by references accessing to $addr_x$ in all references from Core 1. Meanwhile, $\frac{L1WAAD_2[addr_x]}{\sum_{y \in x'\ set} L1AAD_2[y]}$ represents the probability of shared write references to $addr_x$ in all references from Core 2. The numerator $L1WAAD_2[addr_x]$ is the number of writings that access $addr_x$ in L1 private cache of Core 2, while the denominator is the total number of the references that accessing same set as $addr_x$ from Core 1. It is obvious that Eq. (7) and Eq. (17) are very similar in the form as well as insight. This is because Eq. (7) quantifies the data sharing of the shared L2 cache including reading and writing, while Eq. (17) only quantifies the shared writing of L1 private cache.

Similar to Eq. (9), the probability $P1_{split-write}$ that reuse epochs with reuse distance $r$ are split by shared write references come from another core can be represented by Eq. (18):

$$P1_{split-write}(r) = 1 - (1 - P1_{same-write})^{\hat{r}-r} \quad (18)$$

In Eq. (18), $\hat{r} - r$ is the number of references inserted by another core between endpoints of the reuse epoch. Assuming that access frequency from the two cores to its private cache remains relatively uniform, we can derive Eq. (19):

$$\hat{r} - r = r \times \frac{L1_access2}{L1_access1} \quad (19)$$

$L1_access1$ and $L1_assess2$ are the access times during the same execution interval by Core 1 and Core 2 to their corresponding private caches. Eq. (18) describes the probability $P1_{split-write}$ that reuse epochs are split by shared write references coming from another core. Considering another condition of a coherence miss is the accessed cache line still remaining in the private cache, the number of coherence misses can be evaluated by Eq. (20):

$$Miss1_{coherence} = \sum_{r=0}^{\infty} \left[ P1_{split-write}(r) \left( \sum_{sd=0}^{L1_{assoc1}-1} RST1[r]\,[sd] \right) \right] \quad (20)$$

In Eq. (20), $\sum_{sd=0}^{L1_{assoc1}-1} RST[r]\,[sd]$ represents the number of references reuse distance is $r$ and stack distance is less than the associativity. These references should be hits in the private cache, but some of them might be coherence misses. For the references with reuse distance $r$, $P1_{split-write}(r)$ is the probability the reuse epochs are split by shared writings. Therefore, the product of $P1_{split-write}(r)$ and $\sum_{sd=0}^{L1_assoc1} RST1[r]\,[sd]$ represents the number of coherence misses with reuse distance $r$. Accumulating all reuse distances from 0 to infinity, we can calculate all the number of coherence misses of Core 1[2].

In the same manner, the number of coherence misses of Core 2 can be estimated by Eq. (21):

$$\begin{cases} P2_{same-write} = \sum_{addr_x \epsilon S} \left(\frac{L1AAD_2[addr_x]}{L2_access_1} \times \frac{L1WAAD_1[addr_x]}{\sum_{y \in x' set} L1AAD_1[y]}\right) \\ P2_{split-write}(r) = 1 - (1 - P2_{same-write})^{\hat{r}-r} \\ \hat{r} - r = r \times \frac{L1_access1}{L1_access2} \\ Miss2_{coherence} = \sum_{r=0}^{\infty} \left[ P2_{split-write}(r) \left( \sum_{sd=0}^{L1_assoc2} RST2[r]\,[sd] \right) \right] \end{cases} \quad (21)$$

Although our derivation is based on a dual-core architecture, it can be easily extended to other multi-core architectures, like we have introduced in Fig. 6.

### C. *The integration of dwonstream and upstream cache models*

Actually, the integration of downstream and upstream models is the data flow between two models. In the introduction of the upstream model, we have simply introduced how to obtain the input parameter of the downstream model. In order to explain more clearly, this section will introduce the input and output relationships between models in detail.

Fig. 3 shows the data flow of our model framework. We not only profile RST, Hit-RDH and L1RDH input to the upstream cache model, but also profile L1AAD and L1WAAD for the coherence miss model. These inputs are all profiled from CPU trace without time-consuming simulations. The upstream model outputs L2RDH to the shared cache model and obtains AAD by updating code in profiling of RST, which is shown in Fig. 9.

All inputs to the model framework are originally from the CPU traces, some of the input parameters are directly obtained from the CPU trace, and other parameters can be derived through the analytical model. Therefore, the model does not depend on the timing simulations of the cache system, which improves the evaluating efficiency of our approach.

---

[2] In this paper, we assume the L1 private caches use LRU replacement policy, which is commonly seen in commercial processors. It is also worth to note that we cut off the reuse distance extraction at 1024, as we mentioned in Section IV.A. Therefore, accumulating reuse distances in Eq.(20) and Eq.(21) actually is from 0 to 1024.

## V. EVALUATION

The validation including two parts: validating the shared cache model independently, validating the integration of the upstream model and the shared cache model.

### A. *Validating the shared cache model*

We validate our shared cache model against gem5 simulations [19] with the PARSEC version 2.1 on a disk image provided by Computer Architecture and Technology Laboratory, Department of Computer Science of the University of Texas at Austin [20]. The disk image contains pre-compiled statically linked Alpha binaries for all the 13 PARSEC 2.1 benchmarks. The *simsmall* input set of the benchmark is selected in our experiment to limit the simulation time. The applications are divided into three phases: an initial serial phase, a parallel phase, and a final serial phase. Considering that our model focuses on data sharing, we merely perform the validation in the parallel phase, which is also called region of interest(ROI). During this evaluation, the input L2RDHs of the shared cache model are obtained from detailed simulations of each core. We compare the L2 cache miss rates with the gem5 results under 8 hardware configurations including dual-core and quad-core architectures. Each core has private L1 caches, and shares L2 cache with the others. The detailed hardware configurations are shown in TABLE I.

Fig. 12. and Fig. 13. show the experimental results of our model for the 8 hardware configurations, in which the Y axis represents the L2 cache miss rate. Considering that our model's error compared to Gem5 simulations includes the error caused by StatStack, we add the purple bars, i.e., Gem5+StatStack, in the figures besides the L2 miss rates derived from StatCC, gem5 simulations and our method. The L2 miss rates of the purple bars are calculated by StatStack fed with L2MRDH profiled in gem5 simulations. Owing to that our work focuses on the construction of L2MRDH, both our error and StatCC's error are calculated by comparing their results to those of 'Gem5+StatStack' to eliminate the influences from StatStack model.

The average absolute errors of our model and StatCC under dual-core and quad-core architectures are shown in TABLE II. According to the figures, we can see that our method is much more accurate than StatCC under three benchmarks, namely *canneal*, *fluidanimate* and *vips*. The average absolute errors of our model and StatCC with these benchmarks are also shown in TABLE II.

TABLE I. MULTI-LEVEL CACHE HARDWARE CONFIGUGRATION

| **Configuration options** | **Configuration parameters** |
|---|---|
| ISA | ALPHA |
| pipeline | 7-stages, out-of-order |
| L1 cache size | 32KB, 64KB |
| L1 cache associativity | 8-way |
| L1 replacement policy | LRU |
| L2 cache size | 1MB,2MB |
| L2 cache associativity | 16-way |
| L2 replacement policy | LRU |
| DRAM size | 4GB |

TABLE II. AVERAGE ABSOLUTE ERRORS OF THE EVALUATION RESULTS FROM OUR MODEL AND STATCC

| | **Dual-core** | **Quad-core** | **Dual-core with intensive data-sharing** | **Quad-core with intensive data-sharing** |
|---|---|---|---|---|
| Proposed Model | 1.2023% | 1.2013% | 2.7852% | 2.8460% |
| StatCC | 1.2564% | 2.7975% | 9.1965% | 10.7199% |

As we can conclude, although the average absolute error of our model of all benchmarks in the dual-core configurations is only slightly lower than that of StatCC, the error difference between these two models under the quad-core environment has been significantly enlarged. Furthermore, if we only consider the influences of aforementioned three data-sharing intensive benchmarks, the accuracy advantage of our model is apparently obvious and the average errors of our model are less than one third of those of StatCC.

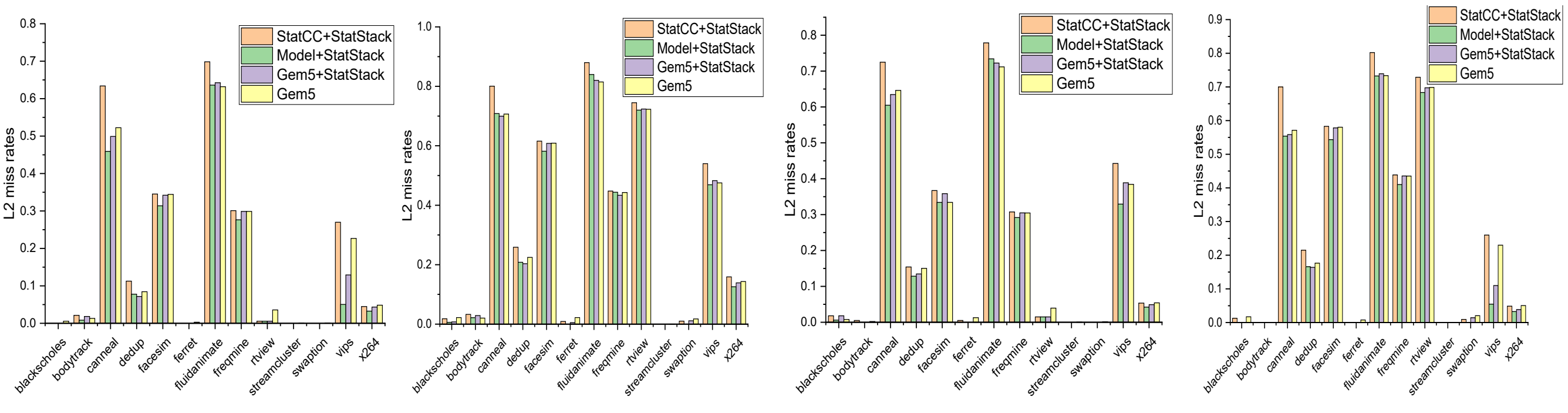

(a) L1 32KB; L2 2MB (b) L1 64KB; L2 1MB (c) L1 32KB; L2 1MB (d) L1 64KB; L2 2MB

Fig. 12. Comparison of L2 miss rates in dual-core architectures

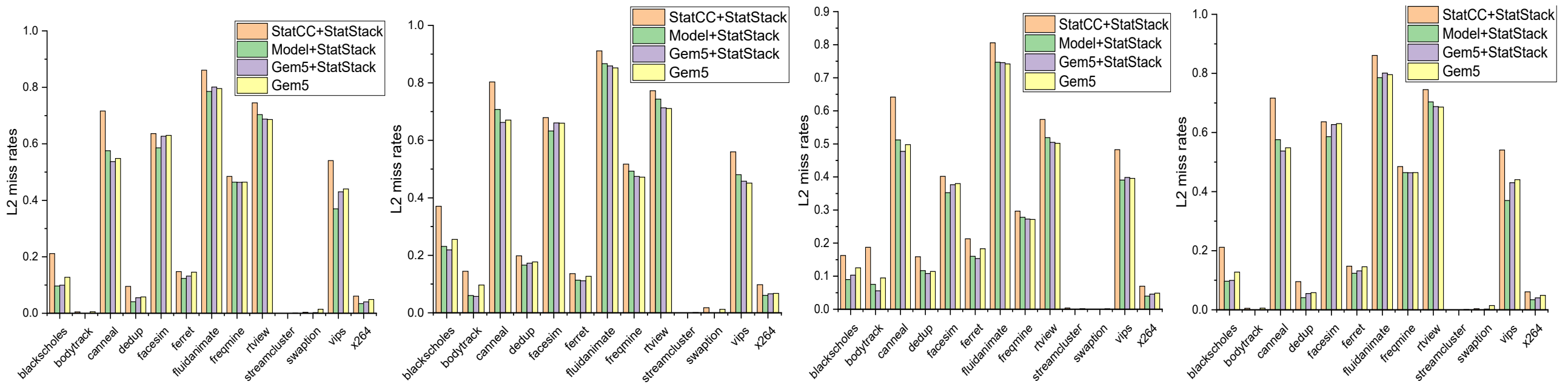

(a) L1 32KB; L2 2MB (b) L1 64KB; L2 1MB (c) L1 32KB; L2 1MB (d) L1 64KB; L2 2MB

Fig. 13. Comparison of L2 miss rates in quad-core architectures

## B. *Validating the integration of upstream cache model and downstream cache model*

The validation of integration of the upstream cache model and the shared cache model includes validating the accuracy of L1 private cache coherence misses and L2 shared cache miss rate under 4 cache configurations shown in Table III.

TABLE III. THE FOUR CACHE CONFIGURATIONS

| | L1 cache size | L1 cache associativity | L2 cache size | L2 cache associativity |
|---|---|---|---|---|
| **Configuration 1** | 128KB | 2 | 1MB | 16 |
| **Configuration 2** | 128KB | 2 | 2MB | 32 |
| **Configuration 3** | 256KB | 4 | 1MB | 16 |
| **Configuration 4** | 256KB | 4 | 2MB | 32 |

It should be noted that the "L1 cache size" in TABLE III is the size of L1 data cache instead of instruction cache. Actually, the size of instruction cache is set as 1MB and 64 associativity to minimize the influences from the L1 instruction cache.

Fig. 14 to Fig. 17 show the results of the coherence miss model compared to results from gem5 in 4 dual-core architectures. We validate the integrated model framework against gem5 simulations [17] with PARSEC 2.1. We choose 10 programs in PARSEC to run on four hardware configurations in a dual-core architecture. The detailed four cache configurations are shown in TABLE III, while other hardware parameters are same as TABLE I.

We use the normalized L1 cache misses to calculate the errors of the coherence miss model. The normalized L1 cache miss is calculated by Eq. (22):

$$Normalize\ L1\ cache\ miss = \frac{misses\ from\ our\ model}{misses\ from\ gem5} \quad (22)$$

The error of coherence miss model is calculated by Eq. (23):

$$Error\ of\ L1\ cache\ miss = \left|\frac{misses\ from\ gem5 - misses\ from\ our\ model}{misses\ from\ gem5}\right| \times 100\% \quad (23)$$

As shown in Fig. 14 to Fig. 17, the normalized miss of most programs before refining by our coherence miss model is near to 1, which means these programs have relatively few coherence misses. Luckily, the results after the application of our coherence model keep the accuracies. For programs with many coherence misses, for example *freqmine* in Fig. 14, cache misses predicted by conventional method show significant errors compared to the simulation results. However, after refining by our coherence miss model, the refined results are close to 1, or the simulation results, which demonstrates the effectiveness of the coherence miss model.

To show the error changes before and after the application of our coherence model more clearly, we give the comparison of the average errors of coherence misses among our model and *uniform model* in [24] of 4 hardware configurations in Fig. 18. According to the comparison results, the average errors of L1 cache misses from our model and *uniform model* show a similar precision. While the time overhead of our model and *uniform model* shows a big difference. Though the overall time consumption contains instruction tracing, profiling and model calculation, the calculation of cache misses is based on the formulas that the time overhead of model calculation is much lower than that of others. Therefore, this paper mainly compares the average time overhead of instruction tracing and profiling in Fig. 19. As shown in the figure, our work takes almost one-tenth of *uniform model*'s time consuming. Since *uniform model* needs to record the RDH/SDH of each shared written address by searching through all the accessed addresses, our work only records the RDH/SDH of each core that reduces much profiling time and memory space.

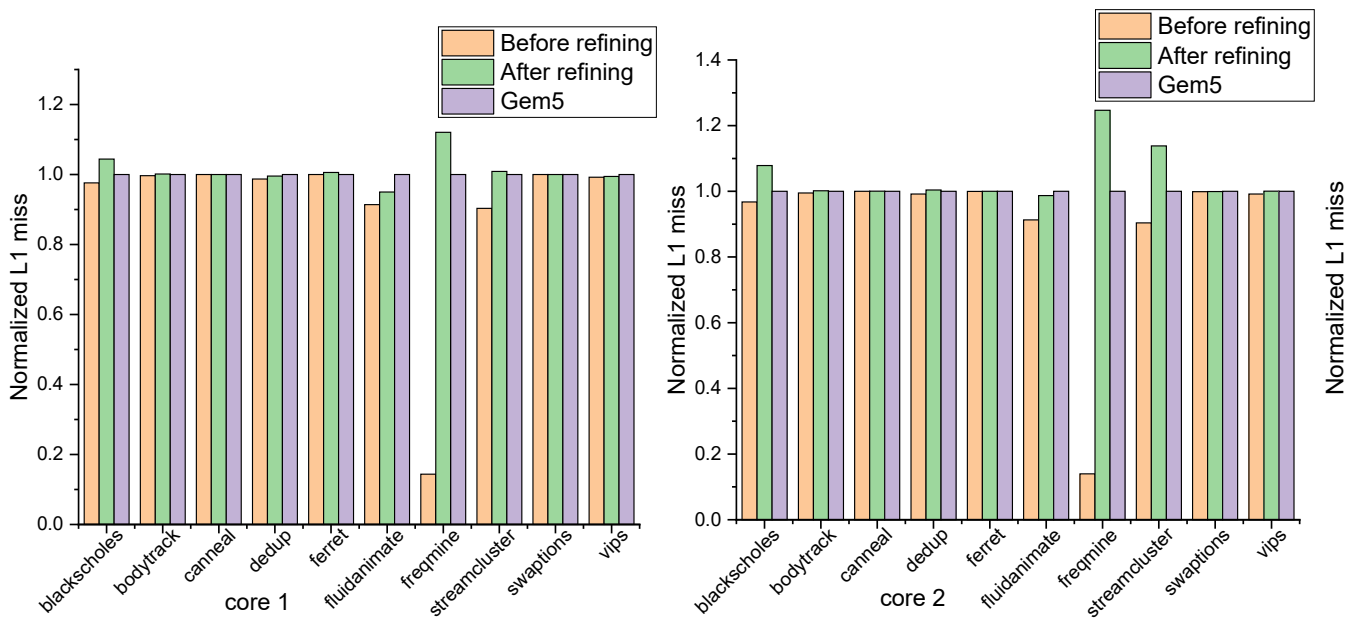

Fig. 14. The results of our coherence miss model under Configuration 1

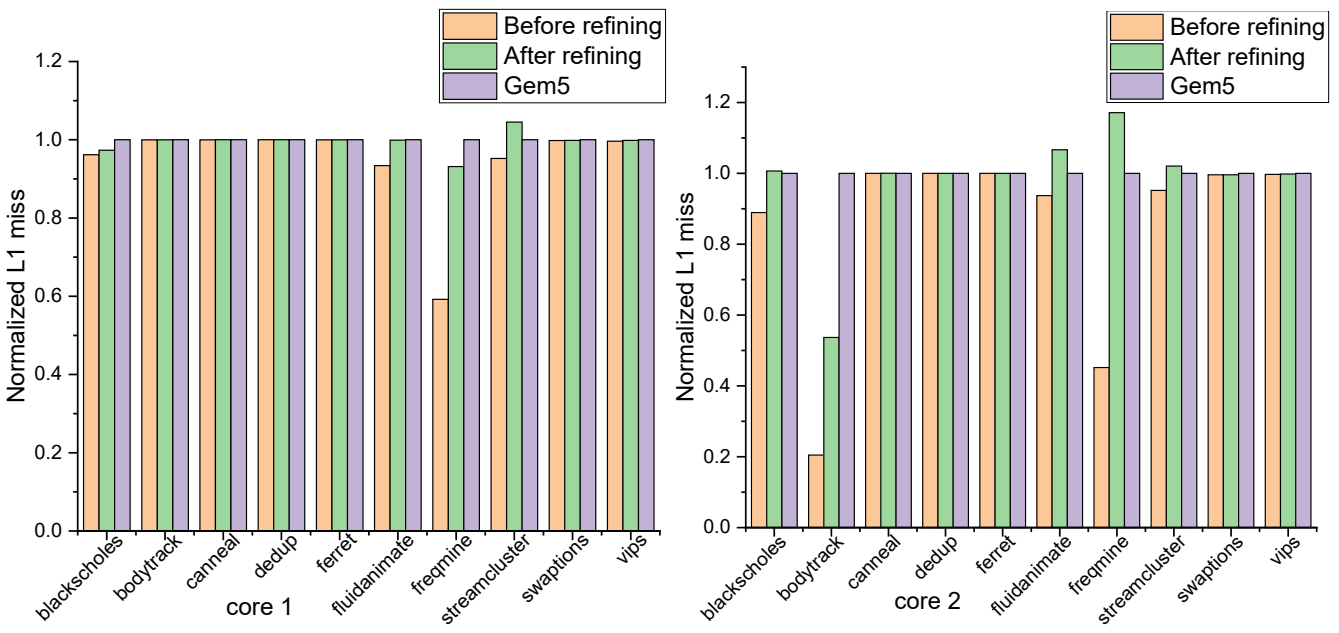

Fig. 15. The results of our coherence miss model under Configuration 2

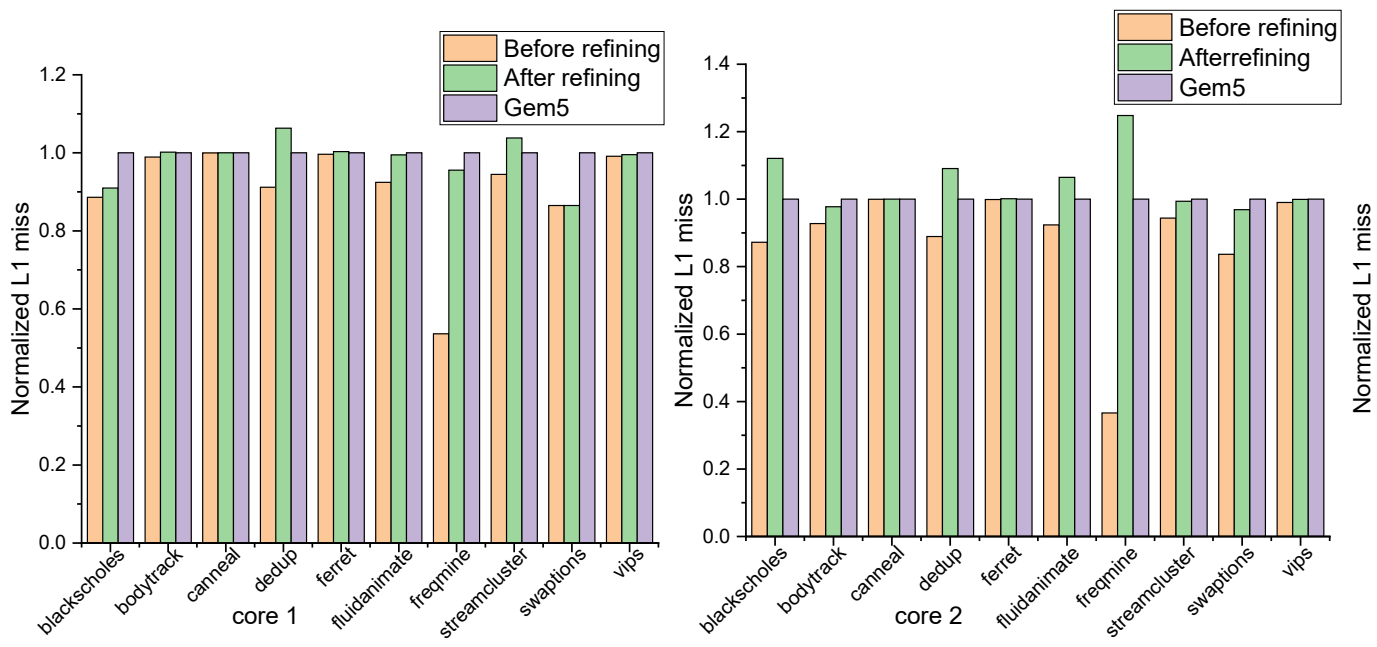

Fig. 16. The results of our coherence miss model under Configuration 3

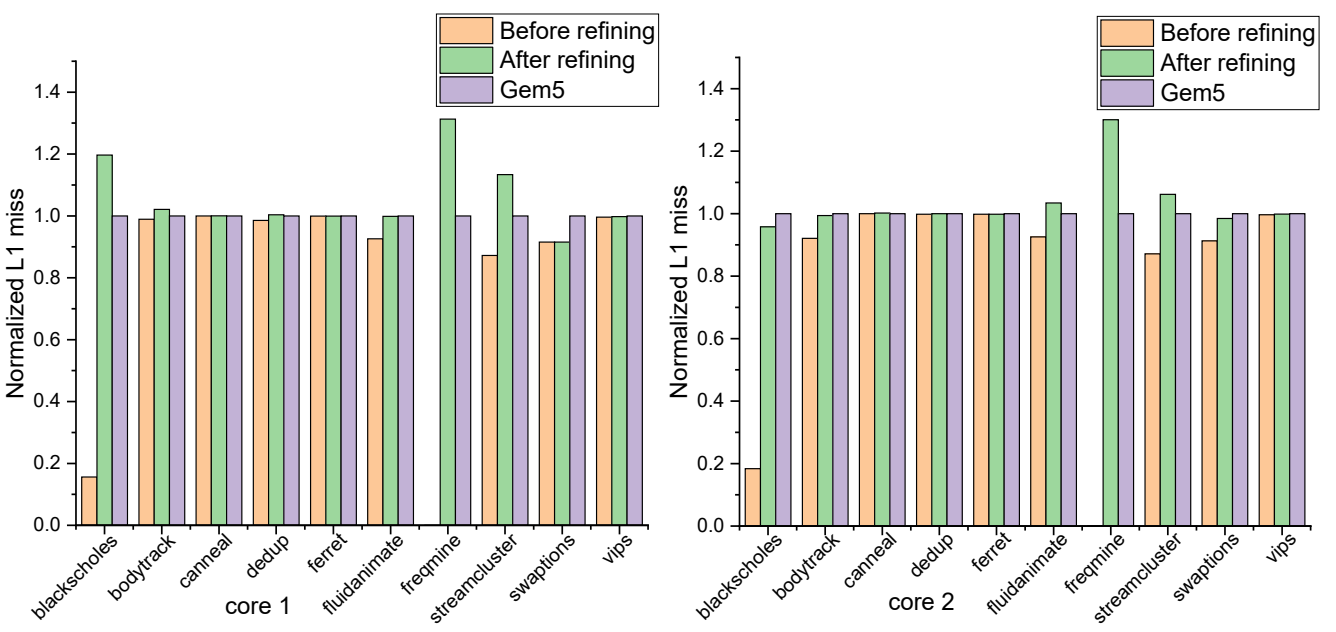

Fig. 17. The results of our coherence miss model under Configuration 4

We also validate the total errors after the integration of the upstream and the shared cache models, the result is shown in Fig. 20. The error is calculated by Eq. (24):

$$error = |MR_{models} - MR_{gem5}| \tag{24}$$

As shown in the Fig. 20, the largest error of the integrated model in the four configurations is about 10% and the average error of the four configurations is 8.03%. As the integrated model framework integrates three analytical models(upstream cache model, shared cache model and StatStack model), each of which contains some ideal assumptions to simplify the modeling, this may enlarge the final errors. For example, all the three models assume that the references are uniform and independent. It is obvious that the assumption could not be reality. We believe the error around 10% is still reasonable and acceptable considering the speed advantage brought by the absence of time-consuming simulations in our evaluation framework.

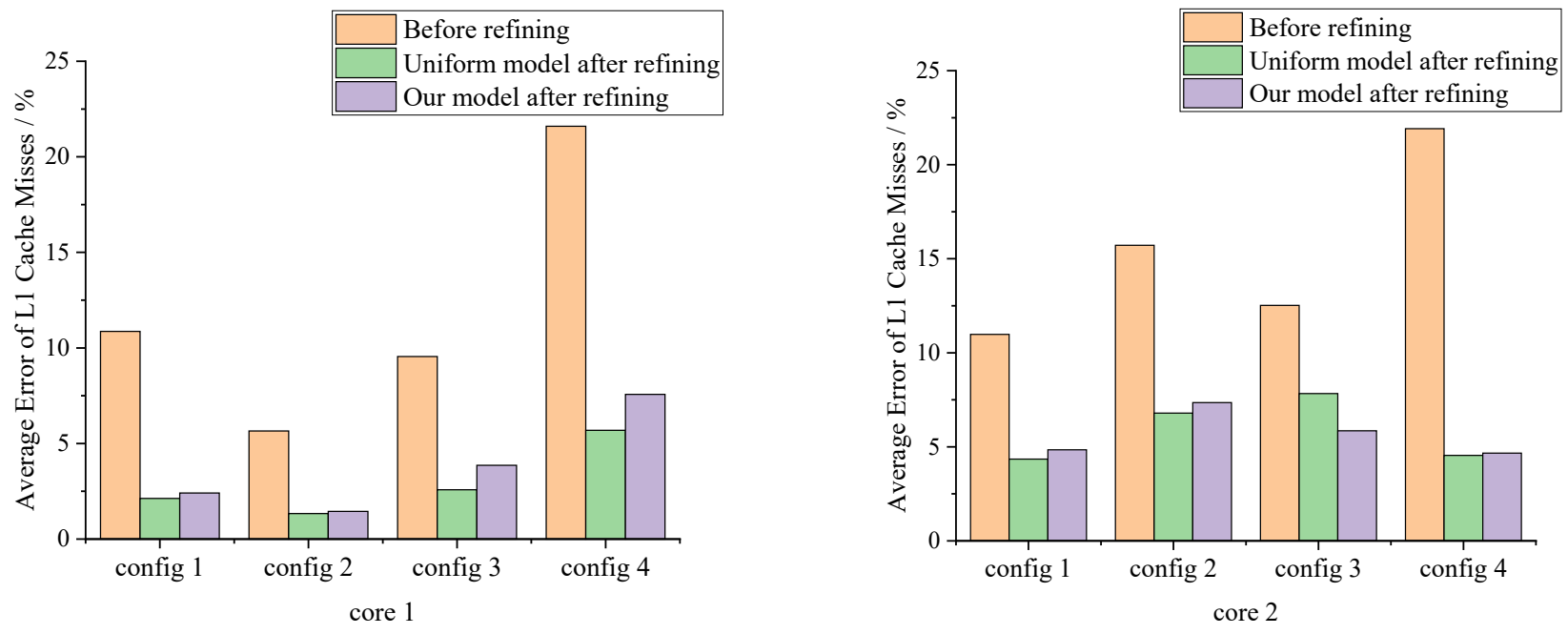

Fig. 18. Average Errors of L1 Cache Misses

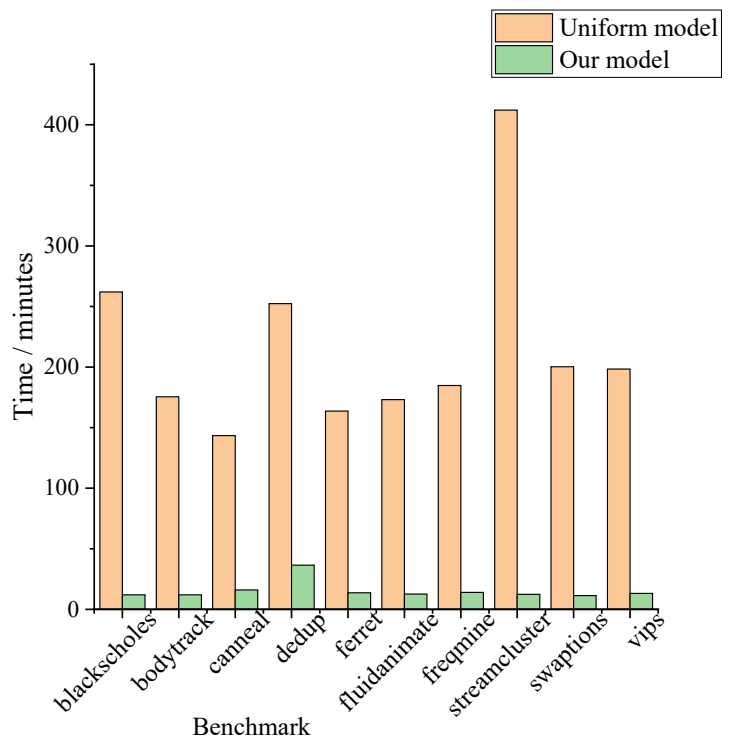

Fig. 19. Comparisons of Estimation Time

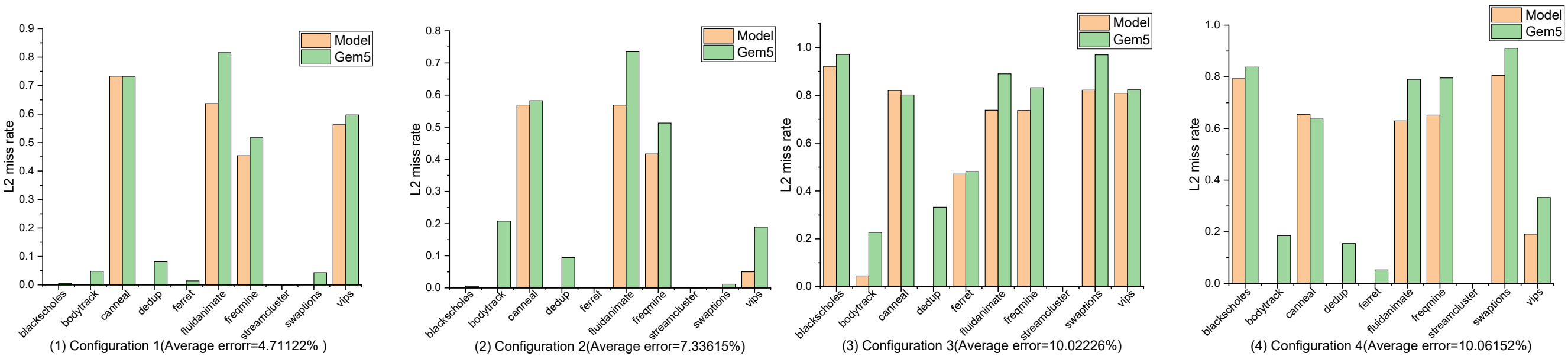

Fig. 20. The result of the integrated model framework in four configurations

## VI. Application of The Integrated Model

Early in the design cycle, architects often use design space exploration(DSE) to determine the choice of processor architecture parameters. For processor architecture design, there are many dimensions that can be selected on the hardware, such as the instruction issue width, ROB size, and the cache capacity. There are many hardware parameters that can be selected for each dimension. These hardware parameters will affect the performance of the processor. For an application that runs on this processor, there is an optimal parameter combination in the design space to achieve performance and energy goals. The purpose of design space exploration is to find such optimized parameter combinations.

For the cache system this paper focuses on, the dimensions include cache capacity, cache associativity, replacement strategy, etc. The optimization goals can be missing rates, power consumption, etc. [19]. By exploring the design space of the cache systems, we can find the optimized cache hardware parameters to achieve the optimized design for the specified target.

Since the design space is usually a combination of parameters of different dimensions, the design space increases exponentially as the dimension increases. If every node in the design space adopts timing simulations to evaluate its performance, the exploration of the entire design space will be extremely time-consuming. Therefore, an analytical model becomes a better choice with the advantage of speed.

In this section, the integrated model is used as an evaluation tool to explore the design space of cache capacity and associativity. In the example of design space exploration, the dual-core processor architecture shown in Fig. 1 is used. The cache system of the dual-core architecture includes private L1 cache and shared L2 cache, because the sharing and competition between cores occurs in the L2 shared cache, and once the reference to the shared cache is a miss, it needs to access the off-chip main memory. The time cost of an off-chip access is 2 to 3 orders of magnitude higher than that of an on-chip memory access. Therefore, in order to ensure the service capability of the entire storage system, the design space exploration will use cache miss of the L2 shared cache as the optimization goal.

Considering different L1 cache hardware parameters results in a different total number of references to the L2 shared cache, it is more reasonable to use the number of misses as an indicator than the miss rate. We use *canneal* in the PARSEC suite as an example to conduct the evaluation experiment.

The DSE will explore four design choices: L1 cache capacity, L1 cache associativity, L2 cache capacity and L2 cache associativity. We choose 57 hardware configurations including partial combinations of L1 cache capacity from 16KB to 256KB and L2 cache capacity from 32KB to 4MB. Except for cache capacity and associativity, all other hardware configurations are the same as Section V. Fig. 21 shows the L2 misses that estimated by the proposed model under 57 cache configurations, in which the abscissa represents the parameters of the cache configuration. For example, "16k-64k-2-8" means 16KB L1 cache and 64KB L2 cache with L1 and L2 associativity of 2 and 8, respectively. The ordinate is the number of L2 shared cache misses, which is estimated by our model in an execution interval with 10 million memory access instructions.

The total cache capacity (the sum of L1 Cache capacity and L2 Cache capacity) on the abscissa shown in Fig. 21 is increasing, so the L2 misses have a decreasing trend. However, the points in this figure are not strictly decreasing in sequence, which means different parameter combinations also have a non-negligible effect on the L2 shared cache misses. According to the results in Fig. 21, if there are no additional constraints such as power consumption and area in the processor design within the search range, selecting the configuration "128k-4M-2-64" can achieve the minimum number of L2 shared cache miss, the point is marked with a purple circle in Fig. 21. If there is a constraint that the cache capacity does not exceed 1M, then "16k-512k-2-64" will be the best parameter selection, marked with a red circle in Fig. 21. According to different design requirements, the optimal configuration within the selectable range can be guided and selected in these configurations. In addition, some data in Fig. 21 can also bring some guiding significance for the architect. For example, under the same capacity of L1 cache and L2 cache, the greater the associativity, the smaller the number of L2 cache misses. It means that for the *canneal* program, under the same capacity constraints, choosing the larger associativity may result in better L2 cache performance.

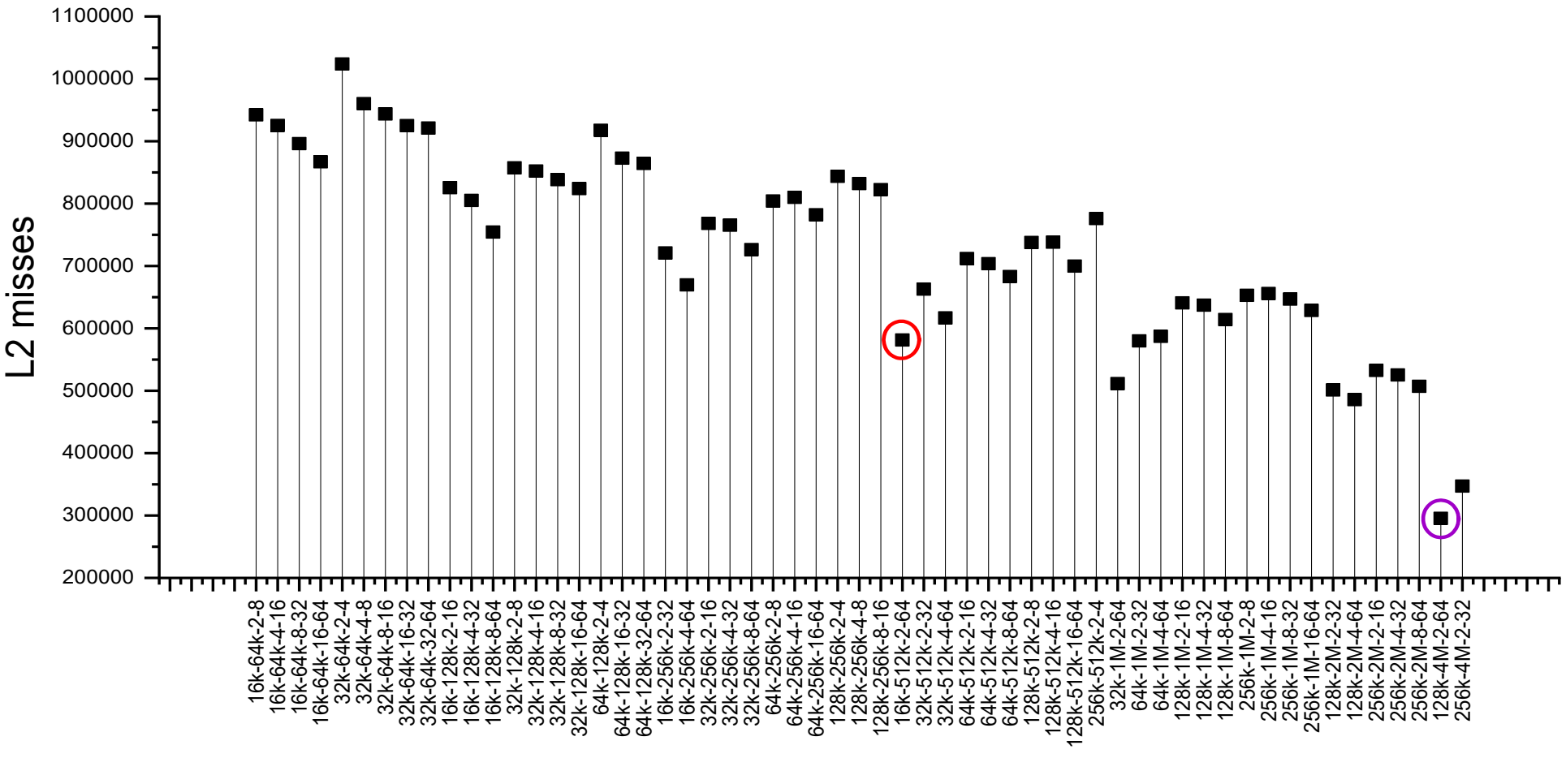

Fig. 21. The L2 miss under 57 kinds of configurations

## VII. Conclusion

In this paper, we have proposed a data-sharing aware shared cache miss rates model for multi-core processors with multi-level cache hierarchies. The Merged Reuse Distance Histograms (MRDH), which represents the RDH of the interleaved access streams from individual cores, is evaluated based on the L2 RDHs and AAD, or Accessing Address Distribution, output by an upstream cache model to avoid the time-consuming full simulations that needed by conventional methods. By a detailed probability derivation, the reuse epoch spit effect, which caused by the data sharing accesses from different cores, can be quantified and used to adjust the MRDH, from which the cache miss rates of the L2 shared cache can be obtained. Moreover, we also refine the upstream cache models with the consideration of multi-core private cache coherent effect. The absolute average errors of 13 benchmarks of the shared cache model are only 1.2% and 1.3% under dual-core and quad-core configurations, respectively. While the average errors of the 3 data-sharing intensive benchmarks are merely one third of those of StatCC. After integrated with the upstream model, the overall average absolute error is 8.03% in 4 hardware configurations. As an example of the proposed model's application, we also evaluate the L2 cache performance under 57 different cache configurations to select the optional design points

## ACKNOWLEDGMENT

This work was supported by the National Natural Science Foundation of China under Grant No. 61974024 and the Provincial Natural Science Foundation of Jiangsu Province under Grant No. BK20181141.